\documentclass[12pt]{article}
\usepackage{amsfonts}
\usepackage{amsmath}
\usepackage{amssymb}
\usepackage{graphicx}
\usepackage{color}
\usepackage[all, knot]{xy}
\usepackage{tikz}
\usepackage{array}
\usepackage{comment}

\usepackage{ulem}

\usepackage[utf8]{inputenc}
\usepackage{epstopdf}
\usepackage[footnotesize]{caption}
\usepackage{amsthm}
\usepackage{enumitem}
\usepackage{mathrsfs}

\usepackage[margin=3cm]{geometry}

\def \be {\begin{equation}}
\def \ee {\end{equation}}
\def \bea {\begin{eqnarray}}
\def \eea {\end{eqnarray}}
\def \nn {\nonumber}

\def \rr {\raise.35ex\hbox{\small $\prime$}\kern-.17em{\mbox{\large $\imath$}}}

\def \dels {\partial\kern-.6em /\kern.1em}
\def \As {{A\kern-.5em / \kern.5em}}
\def \Ds {D\kern-.7em / \kern.5em}

\def \dag {\dagger}

\def \ks {k\kern-.5em /}
\def \ls {l\kern-.5em /}

\newcommand{\ci}[1]{}
\newcommand{\ba}{\begin{eqnarray}}
\newcommand{\ea}{\end{eqnarray}}
\newcommand{\bal}{\begin{align}}
\newcommand{\eal}{\end{align}}
\newcommand{\bay}[1]{\left(\begin{array}{#1}}
\newcommand{\eay}{\end{array}\right)}

\newcommand{\hide}[1]{}

\newlist{axioms}{enumerate}{2}
\setlist[axioms,1]{label=\textbf{A\arabic{axiomsi}.}, ref=A\arabic{axiomsi}}
\setlist[axioms,2]{label=\textbf{A\arabic{axiomsi}\rlap{\myEnumCounter{axiomsii}}.},%
                   ref=A\arabic{axiomsi}\myEnumCounter{axiomsii},%
                   align=parleft,%
                   leftmargin=0em,%
                   itemsep=1.4ex,%
                   before={\stepcounter{axiomsi}}}

  \usetikzlibrary{decorations.markings}

\begin{document}
\begin{titlepage}
%\begin{flushright}
%NORDITA 2019-102 %\\
%October,  2019
%\end{flushright}

\begin{center}

\textbf{\LARGE
Correlated Disorder in the SYK$_{2}$ model
\vskip.3cm
}
\vskip .5in
{\large
Pak Hang Chris Lau$^{a}$ \footnote{e-mail address: phcl2@mit.edu}, Chen-Te Ma$^{b,c,d,e,f}$ \footnote{e-mail address: yefgst@gmail.com},\\ 
Jeff Murugan$^{e}$ \footnote{e-mail address: jeff.murugan@uct.ac.za}, and Masaki Tezuka$^g$ \footnote{e-mail address: tezuka@scphys.kyoto-u.ac.jp} 
\\
\vskip 1mm
}
{\sl
$^a$
National Center for Theoretical Sciences, National Tsing-Hua University,\\
Hsinchu 30013, Taiwan, R.O.C.
\\
$^b$
Guangdong Provincial Key Laboratory of Nuclear Science,\\
 Institute of Quantum Matter,
South China Normal University, Guangzhou 510006, Guangdong, China.
\\
$^c$
 School of Physics and Telecommunication Engineering,\\
 South China Normal University, Guangzhou 510006, Guangdong, China.
\\
$^d$
Guangdong-Hong Kong Joint Laboratory of Quantum Matter,\\
 Southern Nuclear Science Computing Center, 
South China Normal University, Guangzhou 510006, Guangdong, China.
\\
$^e$
The Laboratory for Quantum Gravity and Strings,\\
Department of Mathematics and Applied Mathematics,
University of Cape Town, Private Bag, Rondebosch 7700, South Africa.
\\
$^f$
Department of Physics and Center for Theoretical Sciences,\\
 National Taiwan University,
  Taipei 10617, Taiwan,
R.O.C.
\\
$^g$
Department of Physics, Kyoto University, Kyoto 606-8502, Japan.
}
\\
\vskip 1mm
\vspace{40pt}
\end{center}
\newpage
\begin{abstract}
We study the SYK$_{2}$ model of $N$ Majorana fermions with random quadratic interactions through a detailed spectral analysis and by coupling the model to 2- and 4-point sources. 
In particular, we define the generalized spectral form factor and  level spacing distribution function by generalizing from the partition function to the generating function. 
For $N=2$, we obtain an exact solution of the generalized spectral form factor. 
It exhibits qualitatively similar behavior to the higher $N$ case with a source term. 
The exact solution helps understand the behavior of the generalized spectral form factor. 
We calculate the generalized level spacing distribution function and the mean value of the adjacent gap ratio defined by the generating function. 
For the SYK$_2$ model with a 4-point source term, we find a Gaussian unitary ensemble behavior in the near-integrable region of the theory, which indicates a transition to chaos. 
This behavior is confirmed by the connected part of the generalized spectral form factor with an unfolded spectrum. 
The departure from this Gaussian random matrix behavior as the relative strength of the source term is increased is consistent with the observation that the 4-point source term alone, without the SYK$_2$ couplings turned on, exhibits an integrable spectrum from the spectral form factor and level spacing distribution function in the large $N$ limit. 
\end{abstract}
\end{titlepage}

\section{Introduction}
\label{sec:1}
\noindent
The Wilsonian picture of Nature as described by various effective theories that capture the physics at different scales is a very appealing one \cite{Ma:2018efs}. Organized by size and speed, the physics of most of the known Universe is captured by one of the four effective descriptions. Quantum mechanics, for example, furnishes a precise description of physics at the atomic scales while everyday, macroscopic scales are well described by Newtonian mechanics. Overall, this view offers us a clear description of physics at more or less all accessible scales. Less clear however, is the transition between effective descriptions. Where does Newtonian physics `end' and General Relativity \cite{Wald:1984rg} `take over'? The transition from quantum to classical mechanics is of particular importance in several areas from quantum computing to systems biology \cite{Bryant:2006}, so it would seem then that a proper understanding of this transition is in order. Suffice it to say that this is a hard problem and, as with all such hard problems, physicists make progress by studying toy models, very special examples like the harmonic oscillator and the hydrogen atom that, while exhibiting some of the physics in question, are sufficiently constrained as to be tractable or even completely solvable. However, since such toy models are far from generic, the lessons that we learn from them are, by definition, of limited value. To broaden the scope of our understanding of interacting quantum systems then, it is important to study more generic features of such systems. One such feature that has seen significant development over the past decade and is the focus of this article is {\it chaos}.
\\

\noindent
Classically, chaos refers to the hypersensitivity of a (nonlinear) dynamical system to small perturbations in its initial conditions and is typically diagnosed by the exponential divergence of trajectories in phase space. Quantum chaos, however, is a far more subtle issue. For one, the wavefunction in a closed system is governed by a linear equation making it difficult to use the dynamical equation for wavefunction evolution to define quantum chaos. Second, the uncertainty principle that is at the heart of quantum systems effectively renders the concept of definite ``phase space trajectories" useless. Consequently, other diagnostic tools are required. One, more natural, such tool is the {\it spectral analysis} of the Hamiltonian operator from which quantum analogs of chaos indicators may be defined \cite{Percival:1973}. This analysis is not without its obstacles. Chief among these is the fact that the spectrum in a quantum system --- effectively the diagonalization of a typically large Hamiltonian matrix --- is notoriously difficult to compute exactly. This difficulty was first circumvented by Berry in a novel application of the Wentzel–Kramers–Brillouin (WKB) method to a class of non-integrable systems \cite{Berry:1977zz}. Although the classical and integrable limits in non-integrable systems do not commute in general, Berry showed that the eigenfunctions of a non-integrable Hamiltonian should exhibit Gaussian statistics and since an integrable system cannot possess such eigenfunctions, level statistics essentially distinguishes between integrable and non-integrable systems, albeit in a rather coarse way \cite{Berry:1977}. This result was, in many ways, the precursor to the eigenstate thermalization hypothesis (ETH) \cite{Srednicki:1994} that established the initial connection between quantum chaos and thermalization \cite{Srednicki:1999}.
\\

\noindent
The calculation of the level spacing distribution function in Sinai's billard problem - probably {\it the} prototypical chaotic system - exhibits a random matrix universality \cite{Bohigas:1983er} and a consequent repulsion between energy levels \cite{Dyson:1962es, Dyson:1962es2} and faster than exponential decay of the probability distribution function of the nearest neighbor gap, further differentiating it from an integrable system. Not only does this then lead to a conjecture that {\it a quantum chaotic system with a classical limit should exhibit random matrix statistics} \cite{Bohigas:1983er}, but it also furnishes novel chaos diagnostic. Indeed, since, from the WKB method, calculating the Fourier transform for the energy difference --- a quantity called the {\it spectral form factor} (SFF) --- in a generic bosonic quantum mechanics also manifests characteristic random matrix statistics \cite{Guhr:1997ve} in a short time \cite{Muller:2004nb}, the SFF proves to be a reliable diagnostic of chaos in quantum many-body systems \cite{Dyer:2016pou,Cotler:2016fpe}.  In particular, the SFF displays a dip-ramp-plateau behavior for random matrix theory that appears to be universal and conjectured to be a unique characteristic of chaotic systems \cite{Brezin-Hikami-1996, Okuyama:2018yep}.
\\

\noindent
The last 5 years have seen a surge of interest in quantum chaos in a relatively new direction, quantum gravity. This interest has been precipitated by a rather remarkable quantum many-body system, the Sachdev-Ye-Kitaev (SYK) model  \cite{Kitaev:2015,Maldacena:2016hyu,Sachdev:1992fk}, a (0+1)-dimensional quantum mechanical model of $N$ Majorana fermions with all-to-all Gaussian random four-Fermi interactions\footnote{To distinguish the original SYK model from the variant of interest to us in this article, we will call the 4-Fermi model or the SYK$_{4}$ model.}. Among its many remarkable properties, the model exhibits an emergent low-energy conformal symmetry, is solvable at strong coupling, is maximally chaotic in the sense of saturating the  Maldacena-Shenker-Stanford (MSS) bound \cite{Maldacena:2015waa} on the leading Lyapunov exponent \cite{Larkin:1969} and perhaps most importantly from the perspective of string theorists, is conjectured to be holographically dual to Jackiw-Teitelboim (JT) gravity on AdS$_{2}$. In other words, the JT/SYK duality furnishes the first known concrete example of a (near)AdS$_{2}$/(near)CFT$_{1}$. On the field theory side of this correspondence, by calculating the level spacing distribution function, it was shown in Ref. \cite{Cotler:2016fpe, Garcia-Garcia:2016mno, Garcia-Garcia:2017pzl, Monteiro:2020buv} that the SYK model develops random matrix level statistics over an exponentially large number of eigenvalues. This behavior is not unexpected in the maximally chaotic SYK$_{4}$ model. Motivated by trying to disambiguate between random and chaotic behavior, in Ref. \cite{Lau:2018kpa} we studied a variant of the conventional SYK model with all-to-all Gaussian two-Majorana interactions, an SYK$_2$ model, and showed that it also displays a similar behavior in its spectral form factor \cite{Lau:2018kpa}, while its level spacing distribution function is Poisson \cite{Nosaka:2018iat}. Since the single-particle problem of the SYK$_2$ model is nothing but anti-symmetric random matrix theory \cite{Dumitriu:2009}, we certainly expect the appearance of single-particle level statistics of the random matrix in such a system.
\\

\noindent
Our primary motivation for this article is to study the effect of correlations in the disordered couplings on the SYK$_2$ model. 
One reason for wanting to do so is to understand if some of the well-known effects of long- and short-range correlations in disordered systems, such as the suppression of Anderson localization, persists for SYK$_q$ models. 
Clearly, such an understanding will have ramifications for, for example, black hole information.
Toward this end, we perturb the SYK$_{2}$ model by 2- and 4-point source terms. If the level statistics of random matrices appear in the perturbation region (where the coefficient of the source term is small), we can compute the generating function from Wick's theorem. Hence this perturbation region is near-integrable. 
In order to carry out our analysis, we need to generalize the partition function to a generating function and subsequently define a generalized spectral form factor and generalized level spacing distribution function that, unlike the standard spectral form factor, encodes the information of the correlations into the level statistics. 
We will only consider the constant coefficient of the source terms in this paper. To summarize our results:
\begin{itemize} 
\item We find that the two Majorana fermion ($N=2$) case has an exact solution to the two-point source term that gives a similar behavior to the higher $N$ case when the coefficient of source term does not vanish. The exact solution provides a novel probe of the dip-ramp-plateau behavior exhibited by the SFF.
\item We then calculate the generalized level spacing distribution function and the mean value of the adjacent gap ratio in the SYK$_2$ model and observe a Gaussian unitary ensemble (GUE) in a perturbation region at a large $N$ from the 4-point source term for the intermediate region of source coupling.  
This GUE behavior is confirmed by the connected part of the generalized spectral form factor with an unfolded spectrum. We also analyze the 4-point source term without any random couplings. The result shows an integrable spectrum in the large-$N$ limit, which confirms that the GUE appears in the intermediate region. A similar result appears in the hydrogen atom in a uniform magnetic field \cite{Friedrich:1989tf}.
\item We study the level statistics of the SYK$_{2}$ model for both the bare and unfolded spectra. The result shows that the unfolded spectrum gives a self-consistent result between the generalized level spacing distribution function and the mean value of the adjacent gap ratio, but that was hidden in the bare spectrum. 
\end{itemize}

\noindent
The rest of this paper is organized as follows: We obtain the exact solution for the generalized spectral form factor of the $N=2$ model and a numerical solution for a larger $N$ in Sec.~\ref{sec:2}. We follow this in Sec.~\ref{sec:3} by a calculation of the generalized level spacing distribution function and the mean value of the adjacent gap ratio. Finally, we discuss and conclude in Sec.~\ref{sec:4}.

\section{Generalized Spectral Form Factor}
\label{sec:2}
\noindent
We will begin by defining the model under consideration. While the SYK model (and its variants) are usually defined in the Hamiltonian formulation, since our focus is on the partition function (and its generalization), it will be more convenient to define it through the path integral. As noted above, SYK$_{2}$ is a quantum mechanical model of $N$ Majorana fermions interacting with random 2-fermi interaction and where the random 2-body couplings are drawn from a Gaussian distribution with zero mean and variance $J^{2}/N$. In what follows for the rest of this section, we will first focus on the case of $N=2$, where we obtain an exact solution for the generalized SFF, followed by a numerical solution for the larger $N$ case. The exact $N=2$ solution exhibits qualitatively similar behavior to the higher $N$ case when the source term is turned on. 

\subsection{Generalized Spectral Form Factor in SYK$_2$ Model}
\noindent
The partition function of the SYK$_2$ model is given by
\bea
Z_{\mathrm{SYK}_2}&=&\int d{\cal J} \int D\psi\ \exp\Bigg\lbrack-\int d\tau\ \Bigg(\frac{1}{2}\sum_{j=1}^N\psi_j\dot{\psi}_j
-i\sum_{1\le i_1<i_2\le N}{\cal J}_{i_1i_2}\psi_{i_1}\psi_{i_2}\Bigg)\Bigg\rbrack
\nn\\
&\times&\exp\Bigg(-\sum_{1\le i_1<i_2\le N}{\cal J}_{i_1i_2}^2\frac{N}{2J^2}\Bigg),
\eea
where ${\cal J}_{j_1j_2}$ is the random coupling constant drawn from the normal distribution ${\cal N}(0,J^2/N)$. 
The normalization of the Majorana fermions is given by $\{\psi_i,\psi_j\}=\delta_{ij}$.
In what follows, we will, without loss of generality, set $J=1$. From this partition function, we can define the $2p$-point generating function of the SYK$_2$ model by coupling it to a source term
\bea
Z_{\mathrm{SYKg}_2} &=&\int d{\cal J}\ \int D\psi\ \exp\Bigg\lbrack-\int d\tau\ \Bigg(\frac{1}{2}\sum_{j=1}^N\psi_j\dot{\psi}_j
-i\sum_{1\le i_1<i_2\le N}{\cal J}_{i_1i_2}\psi_{i_1}\psi_{i_2}
\nn\\
&&-i^p\sum_{1\le i_1<i_2<\cdots<i_{2p-1}<i_{2p}\le N}K_{i_1i_2\cdots i_{2p-1}i_{2p}}\psi_{i_1}\psi_{i_2}\cdots\psi_{i_{2p-1}}\psi_{i_{2p}}\Bigg)\Bigg\rbrack
\nn\\
&\times&\exp\Bigg(-\sum_{1\le i_1<i_2\le N}{\cal J}_{i_1i_2}^2\frac{N}{2J^2}\Bigg),
\eea
where $K_{i_1\cdots i_{2p}}$ is the strength of the source. The Hamiltonian for a given set of $\{{\cal J}_{i_1i_2}\}$ can be extracted from the $Z_{\mathrm{SYKg}_2}$ and is given by
\bea
H_{\mathrm{SYKg}_2}
&=&i\sum_{1\le i_1<i_2\le N}{\cal J}_{i_1i_2}\psi_{i_1}\psi_{i_2}
\nn\\
&&+i^p\sum_{1\le i_1<i_2<\cdots<i_{2p-1}<i_{2p}\le N}K_{i_1i_2\cdots i_{2p-1}i_{2p}}\psi_{i_1}\psi_{i_2}\cdots\psi_{i_{2p-1}}\psi_{i_{2p}}.
\label{eqn:HSYKg2}
\eea
With this Hamiltonian containing the contribution from the source term, one can generalize the definition of both the annealed and quenched spectral form factors by replacing $Z_{\mathrm{SYK}_{2}}$ by $Z_{\mathrm{SYKg}_{2}}$ in their respective definitions so that
\bea
g_{\mathrm{ann}}(t, \beta)\equiv\frac{\left\langle|Z_{\mathrm{SYKg}_2}(\beta, t)|^2\right\rangle}{\left\langle|Z_{\mathrm{SYKg}_2}(\beta, 0)|^2\right\rangle} \,;
\qquad
g_{\mathrm{que}}(t, \beta)\equiv\left\langle \left|\frac{Z_{\mathrm{SYKg}_2}(\beta, t)}{Z_{\mathrm{SYKg}_2}(\beta,0)}\right|^2 \right\rangle \,,
\eea
where $\langle {\cal O}\rangle$ is the expectation value of the operator ${\cal O}$ averaged over the random parameters in the system.
The partition function at finite temperature is given by
\bea
Z_{\mathrm{SYKg}_2}(\beta, t)\equiv \mathrm{Tr}\big(e^{-(\beta-it)H_{\mathrm{SYKg}_2}}\big)\,,
\eea
where $\beta$ is the inverse temperature, and $t$ is the time. In this paper, we will only consider the infinite temperature case in which $\beta=0$ and the annealed and quenched spectral form factors coincide. We will denote them both by $g$, where
\bea
g_{\mathrm{ann}}(t, \beta=0)=g_{\mathrm{que}}(t, \beta=0)\equiv g(t, \beta=0).
\eea  
For simplicity, we choose the strength of the source to be
\bea
K_{i_1i_2\cdots i_{2p-1}i_{2p}}=\alpha_{2p}\epsilon_{i_1i_2\cdots i_{2p-1}i_{2p}}, \label{eqn:source}
\eea
where $\alpha_{2p}$ is a constant, and our convention for the Levi-Civita symbol $\epsilon_{i_1i_2\cdots i_{2p-1}i_{2p}}$ is
\bea
\epsilon_{i_1i_2\cdots i_{2p-1}i_{2p}}=1, \qquad i_1<i_2<\cdots i_{2p-1}<i_{2p}.
\label{eqn:LeviCivita}
\eea

\subsection{$N=2$ Fermions}
\begin{figure}
\begin{centering}
\includegraphics[width=1.\textwidth]{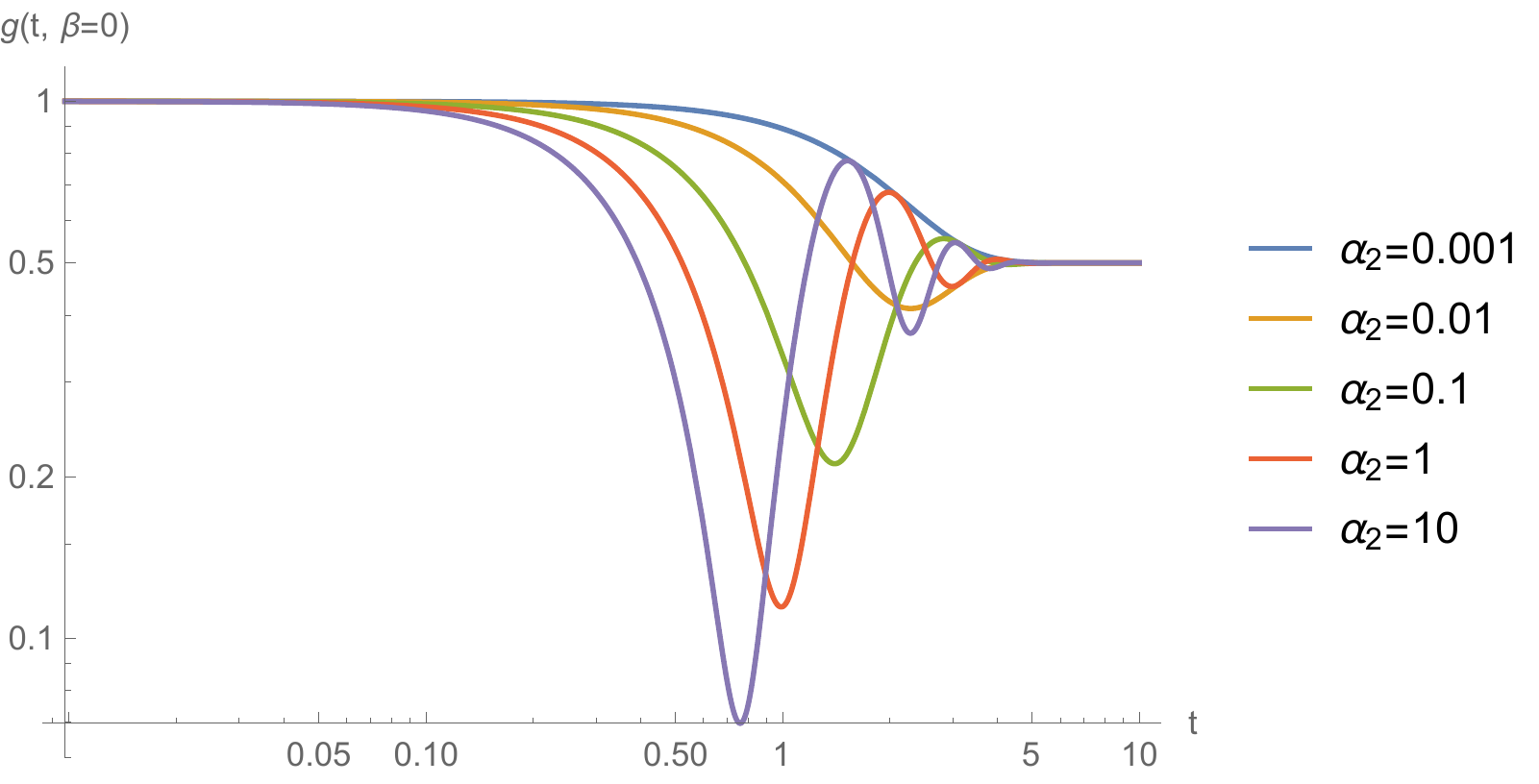}
\end{centering}
\caption{We choose the coefficient of the source term $\alpha_2=0.001, 0.01, 0.1, 1, 10$ to calculate the generalized spectral form factor $g(t, \beta=0)$. When $\alpha_2$ increases, the dip-ramp-plateau behavior appears.
}
\label{SFFN=2.pdf}
\end{figure}

\noindent
In the special case of two fermions interacting with a random quadratic interaction and a 2-point source term, the Hamiltonian becomes:
\bea
H_{\mathrm{SYKg}_2}=i{\cal J}_{12}\psi_{1}\psi_{2}+iK_{12}\psi_{1}\psi_{2}=i({\cal J}_{12}+\alpha_2\epsilon_{12})\psi_1\psi_2,
\eea
where the ${\cal J}_{12}$ are drawn from a Gaussian distribution of vanishing mean and variance $1/2$. In order to compute the associated generating function, it will prove convenient to recast the Hamiltonian in terms of spin variables. To realize this, we use the Jordan-Wigner transformation:
\bea
\psi_1=\frac{\sigma_x}{\sqrt{2}}; \qquad \psi_2=\frac{\sigma_y}{\sqrt{2}}.
\eea
 Our conventions for the Pauli matrices are as follows: They satisfy the usual SU(2) Lie algebra,
	\bea
	\sigma_x\sigma_y=i\sigma_z; \qquad \sigma_y\sigma_z=i\sigma_x; \qquad \sigma_z\sigma_x=i\sigma_y, \nonumber
	\eea
        and their action on the basis vectors $\vert0\rangle = (1,0)^T$ and $\vert1\rangle = (0,1)^T$ is given by
        $\sigma_x\vert0\rangle = \vert1\rangle, \sigma_x\vert1\rangle = \vert0\rangle, \sigma_y\vert0\rangle = i\vert1\rangle, \sigma_y\vert1\rangle = -i\vert0\rangle,\sigma_z\vert0\rangle = \vert0\rangle, \sigma_z\vert1\rangle = -\vert1\rangle$.
The Hamiltonian becomes 
\bea
H_{\mathrm{SYKg}_2}=-\frac{1}{2}({\cal J}_{12}+\alpha_2)\sigma_z.
\eea
% For completeness, we note that the Pauli matrices satisfy the following properties in Appendix \ref{appa}.
In this form, the eigenvalues of the $H_{\mathrm{SYKg}_2}$ are easily read off as $\pm({\cal J}_{12}+\alpha_2)/2$, with  corresponding eigenvectors $|1\rangle$ and $|0\rangle$. The generalized annealed spectral form factor is given by:
\bea
g_{\mathrm{ann}}(t, \beta)&=&\frac{\langle |Z_{\mathrm{SYKg}_2}(\beta, t)|^2\rangle}{\langle |Z_{\mathrm{SYKg}_2}(\beta, 0)|^2\rangle}
\nn\\
&=&
\frac{e^{-\frac{t^2}{4}}\bigg\lbrack e^{\frac{1}{4}(\beta^2-4\alpha_2\beta+t^2)}\bigg(1+e^{2\alpha_2\beta}\bigg)+2\cos(\alpha_2 t)\bigg\rbrack}
{e^{\frac{1}{4}(\beta^2-4\alpha_2\beta)}\bigg(1+e^{2\alpha_2\beta}\bigg)+2} \,,
\eea
where
\bea
&&\langle |Z_{\mathrm{SYKg}_2}(\beta, t)|^2\rangle
\nn\\
&=&\int_{-\infty}^{\infty} d{\cal J}_{12}\ \bigg(e^{-\beta({\cal J}_{12}+\alpha_2)}+e^{\beta({\cal J}_{12}+\alpha_2)}+2\cos\big(({\cal J}_{12}+\alpha_2)t\big)\bigg)e^{-{\cal J}_{12}^2}
\nn\\
&=&\sqrt{\pi}\cdot e^{\frac{\beta^2}{4}-\alpha_2\beta}
+\sqrt{\pi}\cdot e^{\frac{\beta^2}{4}+\alpha_2\beta}
+\sqrt{\pi}\cdot e^{-\frac{t^2}{4}+i\alpha_2 t}
+\sqrt{\pi}\cdot e^{-\frac{t^2}{4}-i\alpha_2 t}
\nn\\
&=&
e^{-\frac{t^2}{4}}\sqrt{\pi}\bigg\lbrack e^{\frac{1}{4}(\beta^2-4\alpha_2\beta+t^2)}\bigg(1+e^{2\alpha_2\beta}\bigg)+2\cos(\alpha_2 t)\bigg\rbrack,
\eea
and in which we have used
\bea
\int_{-\infty}^{\infty} dx\ e^{-ax^2+bx+c}=\sqrt{\frac{\pi}{a}}\cdot e^{\frac{b^2}{4a}+c}\, .
\eea
At $\beta=0$, we obtain a simple solution:
\bea
  g(t)\equiv g_{\mathrm{ann}}(t, \beta=0)=\frac{1+e^{-\frac{t^2}{4}}\cdot\cos(\alpha_2 t)}
{2}.
\eea
Even though this is clearly an oversimplification, displaying none of the wild fluctuations usually exhibited by the SFF prior to ensemble averaging, its functional simplicity allows us to extract some useful (and hopefully universal) features. We note the following:
\begin{itemize}
  \item  The value of the SFF at the two extremes of the half-line are, as expected, 
  \bea
  g(t\to0) = 1\ \ \mathrm{and}\ \ g(t\to\infty) = \frac{1}{2}.
  \eea
  \item When the source term is turned off, 
  \bea
  g(t) = \frac{1+\exp\big(-\frac{t^{2}}{4}\big)}{2}
  \eea
   decreases monotonically from 1 at $t=0$ to $1/2$ at $t\to\infty$, displaying no dip or ramp-like behavior.
  \item When the 2-point source term is turned on, for fixed $\alpha_{2}$, the generalized SFF has turning points whenever 
  \bea
  \alpha_{2} = \cot(\alpha_{2}t)
  \eea
   asymptoting to $1/2$ at late times. The larger the coupling to the source term, the more oscillations that $g(t)$ undergoes before the plateau. Clearly, any dip-ramp-plateau-type behavior arises from the oscillating function $\cos(\alpha_2 t)$.
\end{itemize}
In Fig. \ref{SFFN=2.pdf}, we plot the generalized spectral form factor with the coefficient of the source term $\alpha_2=0.001, 0.01, 0.1, 1, 10$. For $\alpha_2=0.01. 0.1, 1, 10$, we find the dip-ramp-plateau like behavior, numerically confirming our observations above.

\subsection{$N>2$ Fermions}
\begin{figure*}
\includegraphics[width=1.\textwidth]{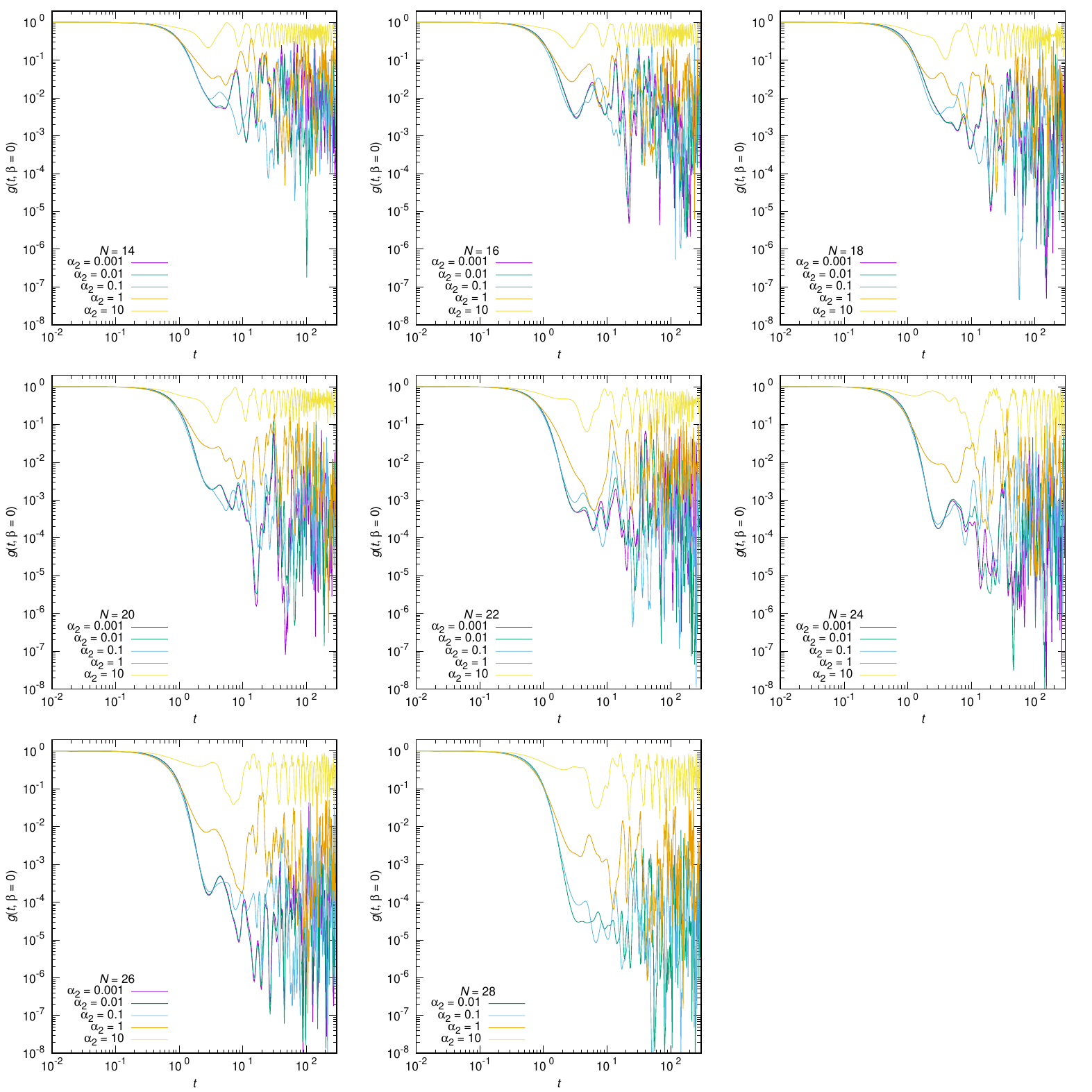}
\caption{The generalized spectral form factor for the SYK$_2$ model + 2-point source term case for $N = 14, 16, 18, 20, 22, 24, 26, 28$ and $\alpha_2=0.001, 0.01, 0.1, 1, 10$. 
\label{SYK2STv1-N14-28-T0001-10F0Gt}}
\end{figure*}

\begin{figure*}
\includegraphics[width=1.\textwidth]{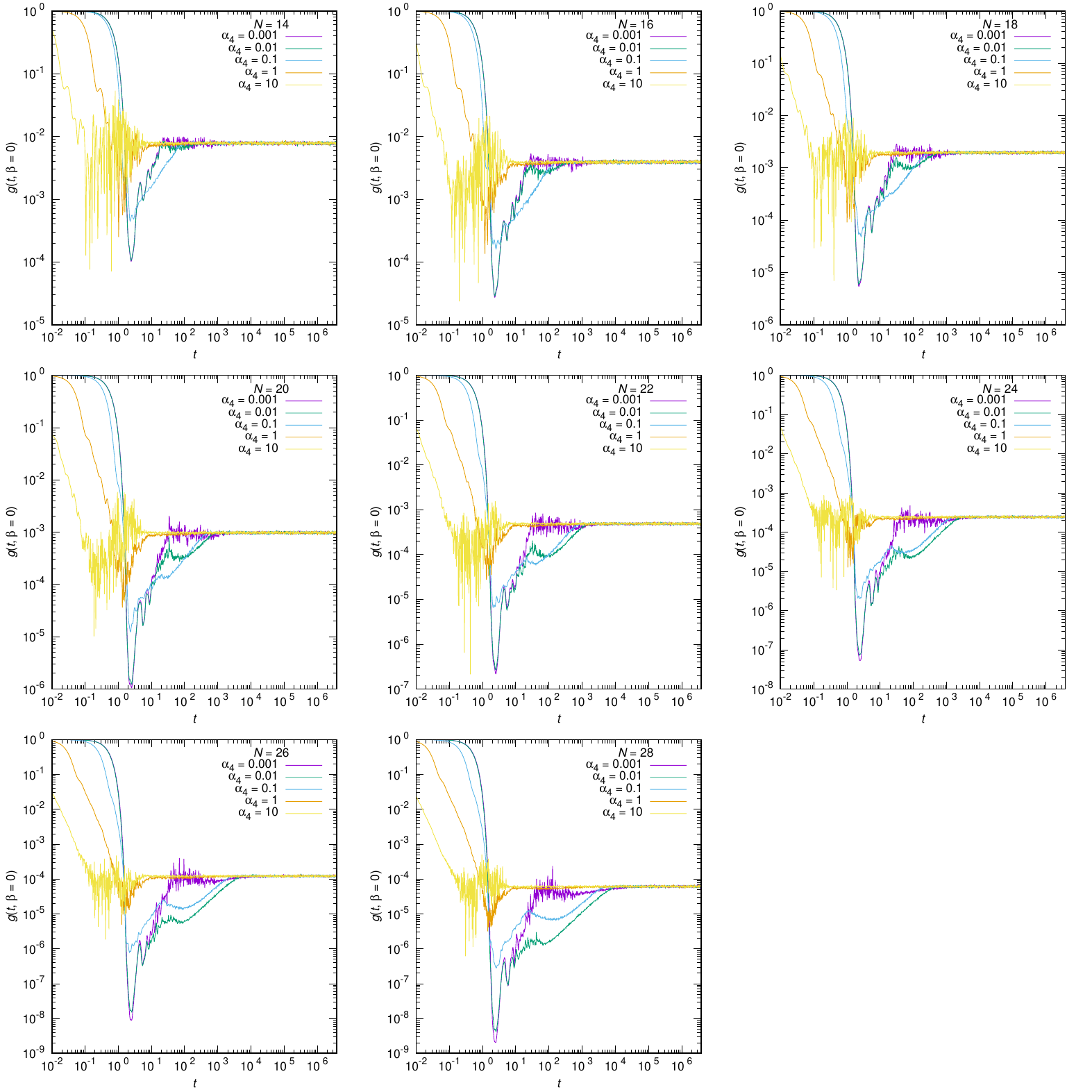}
\caption{The generalized spectral form factor for the SYK$_2$ model + 4-point source term case for $N = 14, 16, 18, 20, 22, 24, 26, 28$ and $\alpha_4=0.001, 0.01, 0.1, 1, 10$.
\label{SYK2STv1-T0F-F-Gt}}
\end{figure*}

\noindent
Unfortunately, for more than two fermions, the situation is analytically intractable, and we have to resort to numerics if the source terms involve more than two fermions. For the rest of the paper, we have used 1000 Gaussian random samples to generate each figure, unless stated otherwise. For the 2-point source terms, once the eigenvalues of the hermitian matrix whose upper triangular part is $i({\cal J}_{i_1i_2}+\alpha_2 \epsilon_{i_1i_2})$ is known, we can evaluate the spectral form factor. In Fig. \ref{SYK2STv1-N14-28-T0001-10F0Gt}, we plot the generalized spectral form factor in the SYK$_2$ model with the 2-point source term for $N = 14, 16, 18, 20, 22, 24, 26, 28$ and $\alpha_2=0.001, 0.01, 0.1, 1, 10$. In Fig. \ref{SYK2STv1-T0F-F-Gt}, we plot the generalized spectral form factor in the SYK$_2$ model with a 4-point source term again for $N = 14, 16, 18, 20, 22, 24, 26, 28$ and $\alpha_4=0.001, 0.01, 0.1, 1, 10$.
\\

\noindent
As can be seen from the plots, the $N>2$ SFF exhibits qualitatively similar behavior to the $N=2$ case with a source term. In this sense, the exact $N=2$ solution is somewhat of a prototype for the larger $N$ SFF. To do better and give a quantitative measure of the difference between the level statistics of random matrix and non-random matrix behaviors, we study the level spacing distribution function in the following section. As noted, the dip-ramp-plateau behavior in the SFF appeared when we turned on a source term in the $N=2$ case. However, since the source coupling is not a random variable, suitable level statistics needs a large $N$. Although the $N=2$ case does not possess a random eigenstate, and the dip-ramp-plateau behavior still appears for some non-vanishing source terms, it alone cannot disambiguate between the randomness of the eigenstates and the dip-ramp-plateau behavior \cite{Lau:2018kpa}. When the source term vanishes, the exact solution shows no ramp behavior as expected.

\section{Random Matrix Behavior}
\label{sec:3}
\noindent
Now we calculate the generalized level spacing distribution function, defined by the generating function, for the 2-point and 4-point source terms. To give a detailed analysis of the level statistics, we also calculate the mean value of the adjacent gap ratio $\langle r\rangle$, where the adjacent gap ratio is defined by
\bea
   r_j\equiv\frac{\mathrm{min}(\delta_j, \delta_{j+1})}{\mathrm{max}(\delta_j, \delta_{j+1})},
\eea
and 
\bea
\delta_j\equiv E_j-E_{j-1},\label{eqn:deltaj}
\eea
where the subscript $j$ denotes the $j$-th eigenvalue of an ordered spectrum ($E_{j-1}<E_j<E_{j+1}$).
 In the case of a 4-point source term, we obtain a Gaussian Unitary Ensemble (GUE) \cite{Dyson:1962es} in the perturbation region and the generalized spectral form factor in the 4-point case shows a clear difference between random matrix and non-random matrix behavior in the ramp region. The random matrix behavior can also be confirmed by the connected part of the generalized spectral form factor of an unfolded spectrum. We analyzed the spectrum for the 4-point source term without a random term with the result suggesting an integrable spectrum in the large-$N$ limit. In summary, then, our model (SYK$_{2}$ + 4-point source term) interpolates between two integrable extremes with a random matrix region in between and for certain values of the couplings.

\subsection{Bare and Unfolded Level Spacing Distribution}
\label{sec:unfold}
\noindent
The level spacing distribution encodes important information about the statistical properties of the system, obtained by collecting data from many samples. The number of samples used in our calculations will be stated along with the results presented below. To normalize the level spacing distribution, we also introduce the mean energy difference between the $(j-1)$-th and $j$-th states averaged over the number of samples:
 \bea
 \langle \delta_j \rangle=\langle E_j-E_{j-1}\rangle= \sum_{k=1}^n \frac{\left(E_j-E_{j-1} \right)_k}{n},
 \eea
where $k$ labels the $k$-th sample, and $n$ is the total number of samples.
For the model we consider, we have $D=2^{N/2-1}$ eigenstates in each parity sector, and we treat the two parity sectors as different samples in analyzing the spectral statistics, thought the number of samples discussed in this paper is the number of the sets of $\{\mathcal{J}_{i_1i_2}\}$ and not the twice thereof.
There are two ways of defining the level spacing distribution; the {\it unfolded} level spacing distribution $P(s)$ uses the unfolded level gaps 
\bea
%s=
\left\{
\frac{\delta_2}{\langle \delta_2\rangle}, \frac{\delta_3}{\langle \delta_3\rangle}, \ldots, \frac{\delta_D}{\langle \delta_D\rangle}
\right\}
\eea
 while the {\it bare} level spacing distribution uses the average of energy differences, 
 \bea
 \overline{\langle \delta\rangle}\equiv \sum_{j=2}^{D}\frac{\langle \delta_j\rangle}{D-1},
 \eea
  to normalize the level spacings 
 
\begin{align}
%  s = 
\left\{\frac{\delta_2}{\overline{\left\langle \delta \right\rangle}} , \cdots , \frac{\delta_{D}}{\overline{\left\langle \delta \right\rangle}}\right\} \,.
  \end{align}
Note that $\delta_j$ carries the sample index $k$ but is omitted for visual clarity. 
Our numerical results show that the unfolded spectral statistics is preferred for a self-consistent study.

\subsection{Number Variance}
\noindent
For the unfolded spectrum, we also calculate the number variance 
\bea
\Sigma^2(L)\equiv\langle N^2(L)\rangle-\langle N(L)\rangle^2,
\eea 
which is the variance of the number of unfolded eigenvalues ($N(L)$) found in an energy range of width $L$ around the center of the energy spectrum.
Here we take the range centered on the average of the unfolded eigenvalues. 
The expectation value of the number of unfolded eigenvalues is $L$ in the limit of a large number of samples and for $L$ sufficiently smaller than the number of eigenenergies.
For the exact Wigner-Dyson ensembles, $\Sigma^2(L)$ is known to depend on $L$ as \cite{Dyson:1962es2}
\[
\Sigma^2(L)\simeq \frac{2}{\pi^2\beta}\log L + (\beta\mathrm{-dependent~constant}),
\]
in which $\beta=1, 2, 4$ for Gaussian orthogonal, unitary, and symplectic ensembles (GOE, GUE, and GSE) respectively.
For the SYK$_4$ model, the Hamiltonian is given by
\bea
H_{\mathrm{SYK}_4}
\equiv-\sum_{1\le i_1<i_2<i_3<i_4\le N}{\cal K}_{i_1i_2i_3i_4}\psi_{i_1}\psi_{i_2}\psi_{i_3}\psi_{i_4},
\eea
in which ${\cal K}_{i_1i_2i_3i_4}$ are independently drawn from a Gaussian distribution of vanishing mean and a constant variance (often chosen to be $6/N^3$),
the eigenvalue statistics show good agreement with the Wigner-Dyson ensembles having the same symmetry, namely
GOE for $N \equiv 0$ (mod 8), GUE for $N \equiv 2,6$ (mod 8), and GSE for $N \equiv 4$ (mod 8) \cite{Cotler:2016fpe}.
On the other hand, an uncorrelated distribution (or a Poisson distribution) has $\Sigma^2(L) = L$.

\subsection{The Properties of the 4-Point Source Term}
\noindent
Before discussing the effect of the source terms, let us first discuss the spectral statistics of the case, where the Hamiltonian consists of the 4-point source alone. As in the below, the spectral statistics does not indicate any spectral correlation. Note that here we do not have random variables, and the spectra analyzed only consists of the two energy spectra corresponding to the two parity sectors each having $D=2^{N/2-1}$ eigenvalues.

\subsubsection{Density of States}
\begin{figure}
%\centering
\includegraphics[width=0.5\textwidth]{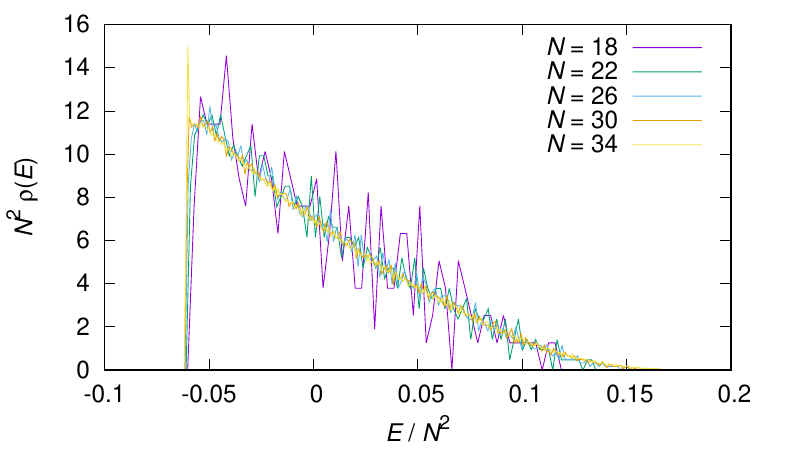}
\includegraphics[width=0.5\textwidth]{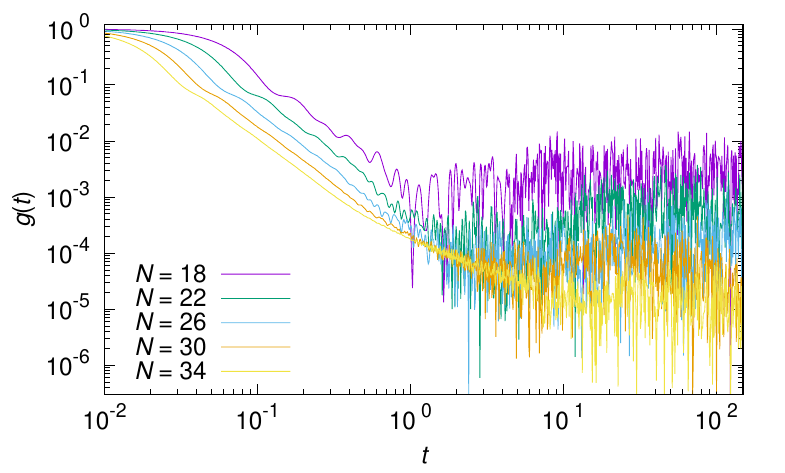}
\caption{Left: The density of states, and right: the spectral form factor for the bare eigenvalues,
for the model \eqref{eqn:HSYKg2} with ${\cal J}_{i_1i_2}=0$, $p=2$, and
$K_{i_1i_2i_3i_4}$ given by Eqs.~\eqref{eqn:source} and \eqref{eqn:LeviCivita} with $\alpha_4=1$.}
\label{fig:source4-dos-gt}
\end{figure}

\noindent
In the left part of Fig.~\ref{fig:source4-dos-gt}, we plot the density of states $\rho(E)$ for the Hamiltonian that consists only of the 4-point source term given by Eqs. \eqref{eqn:source} and \eqref{eqn:LeviCivita}.
The energy distribution scales well with $N^2$, reflecting the fact that there are ${\cal O}(N^4)$ terms in the Hamiltonian
\footnote{
As $N$ is increased, the dependence of $\rho(E)$ on $E$ can be better approximated by $\propto(C-E/N^2)^{\eta}$, with $C\sim 0.163$ and $\eta\sim 1.67 (\sim 5/3)$.}.

\subsubsection{Spectral Form Factor}
\noindent
In the right part of Fig.~\ref{fig:source4-dos-gt}, we plot the spectral form factor for the energy spectrum without an unfolding.
As $N$ is increased, the slope exponentially extends to a smaller value of $g(t)$, which is not followed by a ramp.

\subsubsection{Level Spacing Distribution}
\begin{figure}
%\centering
\includegraphics[width=1.\textwidth]{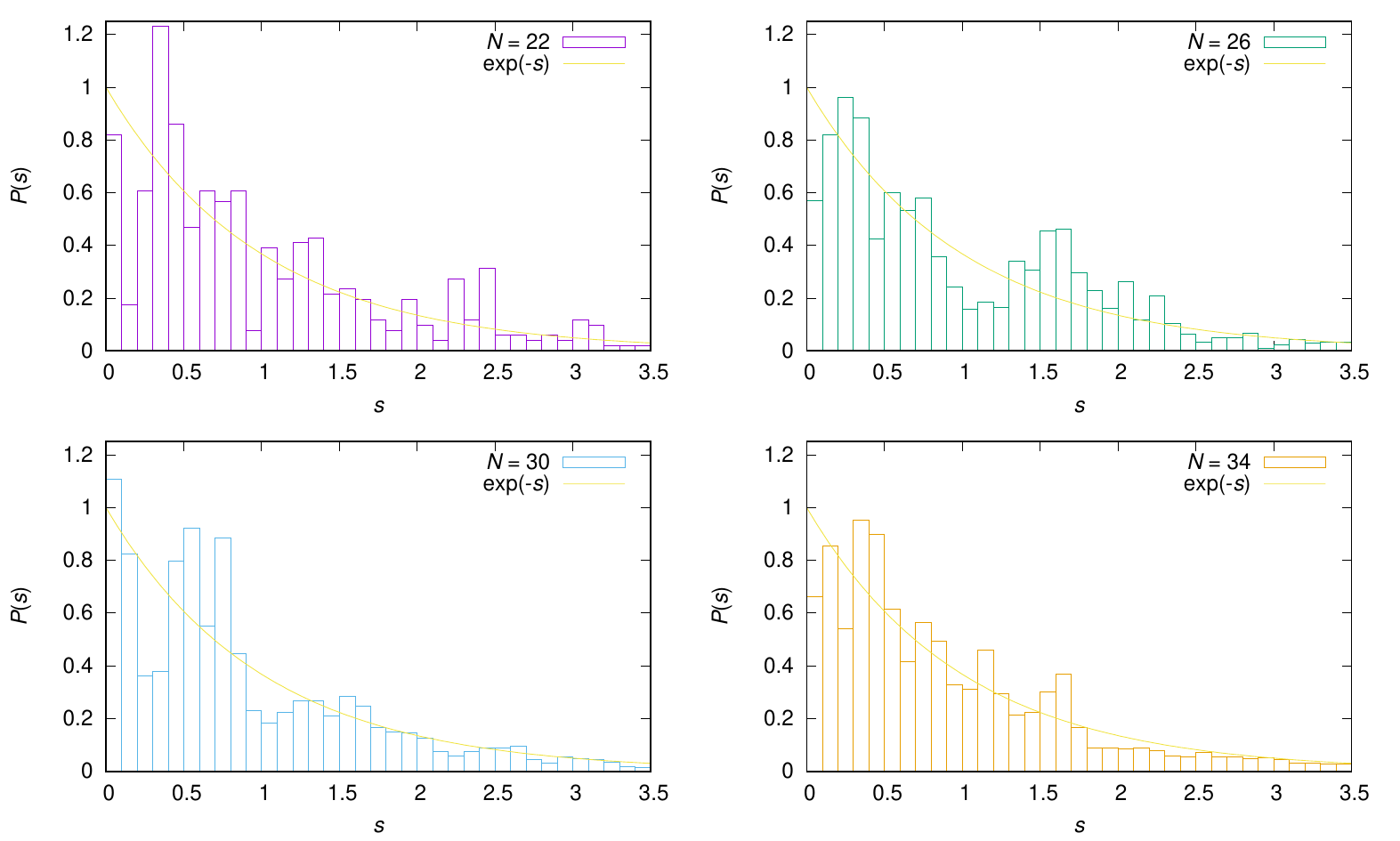}
\caption{
The level spacing distribution for unfolded eigenvalues in the model \eqref{eqn:HSYKg2} with ${\cal J}_{i_1i_2}=0$, $p=2$, and
$K_{i_1i_2i_3i_4}$, given by Eqs.~\eqref{eqn:source} and \eqref{eqn:LeviCivita}.}
\label{fig:source4-unf-Ps}
\end{figure}

\noindent
For each $N$, we fit the density of states $\rho(E)$ using a tenth-order polynomial $\tilde{\rho}(E)$ of the energy $E$,
and use $\tilde{\rho}(E)$ to unfold the central 50\% of the spectrum.  
The distribution of the unfolded level spacing $s$ is shown in Fig.~\ref{fig:source4-unf-Ps}.
The results indicate a gradual approach to the Poisson distribution $\exp(-s)$ as $N$ is increased.

\subsection{SYK$_2$ Model with 2-Point Source Term}
\noindent
We note that the Hamiltonian consisting only of the 2-point source term is solvable.
For a skew-symmetric matrix $G$ with 
\bea
G_{i_2i_1} = -G_{i_1i_2} = i\epsilon_{i_1i_2},
\eea
\begin{equation}
H=\alpha_2\sum_{1\leq i_1<i_2\leq N}G_{i_1i_2}\psi_{i_1}\psi_{i_2}
=\frac{\alpha_2}{2}\sum_{1\leq i_1,i_2\leq N}G_{i_1i_2}\psi_{i_1}\psi_{i_2}
\end{equation}
can be block-diagonalized as
\begin{equation}
H=\alpha_2\sum_{1\leq j\leq \frac{N}{2}}i m_j\tilde{\psi}_{2j-1}\tilde{\psi}_{2j},
\end{equation}
in which $\pm im_j$ are the eigenvalues of $G$ and $\{\tilde{\psi}\}$ are linear combinations of $\{\psi\}$
satisfying 
\bea
\{\tilde{\psi}_j,\tilde{\psi}_k\}=\delta_{jk}.
\eea
We can introduce $N/2$ complex fermions, 
\bea
c_j=\frac{\tilde{\psi}_{2j-1}+i\tilde{\psi}_{2j}}{\sqrt{2}},
\eea
so that 
\bea
\tilde{\psi}_{2j-1}=\frac{c_j+c_j^\dag}{\sqrt{2}}\qquad \mathrm{and}\qquad \tilde{\psi}_{2j}=\frac{c_j-c_j^\dag}{\sqrt{2}i}.
\eea
By defining 
\bea
n_j \equiv c_j^\dag c_j,
\eea
 we have 
 \bea
 i\tilde{\psi}_{2j-1}\tilde{\psi}_{2j}=\frac{2n_j-1}{2},
 \eea
then
\begin{equation}
H = \alpha_2 \sum_{1\leq j\leq \frac{N}{2}}m_j\frac{2n_j-1}{2}.
\end{equation}

\noindent
For the case only with the 2-point source term, we can explicitly obtain $\{m_j\}$ as follows:
Let us first obtain the eigenvalues of an $N$-dimensional Hermitian matrix
\begin{equation}
M = i\begin{pmatrix}
0 & 1 & 1 & \cdots & 1\\
-1& 0 & 1 & \cdots & 1\\
-1& -1& 0 & \cdots & 1\\
\vdots & \vdots & \vdots & \ddots & \vdots\\
-1& -1& -1& \cdots & 0
\end{pmatrix}.
\end{equation}
For $c=\pm1,\pm3,\ldots,\pm(N-1)$, $\phi=\exp(-ic\pi/N)$ satisfies $\phi^N=-1$.
An $N$-dimensional vector
\begin{equation}
\Psi = \begin{pmatrix}
\phi,\phi^2,\phi^3,\ldots,\phi^N
\end{pmatrix}^\mathrm{T},
\end{equation}
is one of the eigenvectors of $M$ because the $k$-th element ($1\leq k\leq N$) of $M\Psi$ is given by:
\begin{align}
(M\Psi)_k &= i\left(-\sum_{l=1}^{k-1}\phi^l+\sum_{l=k+1}^{N}\phi^l\right)
=i\left(
-\frac{\phi^k-\phi}{\phi-1}+\frac{\phi^{N+1}-\phi^{k+1}}{\phi-1}\right)\nonumber\\
&=i\frac{\phi(\phi^N+1)-\phi^k(\phi+1)}{\phi-1}=\left(\cot\frac{c\pi}{2N}\right)\phi^k.
\end{align}
Therefore, we can choose 
\bea
m_j=\cot\frac{(2j-1)\pi}{2N}>0,\ j=1,2, \cdots, \frac{N}{2},
\eea
and the many-body eigenenergies are
\begin{equation}
\epsilon_{(\sum_{j=1}^N 2^{p_j} p_j)}=\frac{\alpha_2}{2}\sum_{j=1}^N(-1)^{p_j}\left(\cot\frac{(2j-1)\pi}{2N}\right)\hspace{2em} (p_j=0,1).
\end{equation}
\\

\begin{figure*}
\includegraphics[width=1.\textwidth]{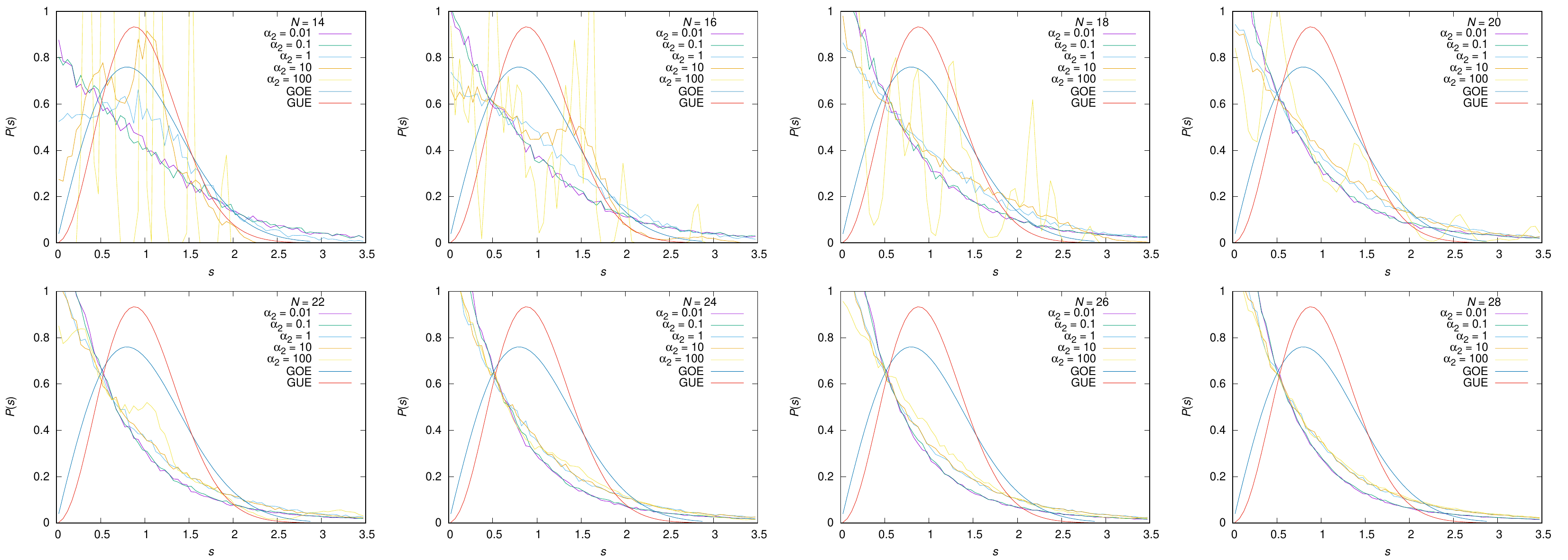}%
\caption{The bare generalized level spacing distribution function for the SYK$_2$ model + 2-point source term case ($\alpha_2=0.01, 0.1, 1, 10, 100$).
Top row: $N=14, 16, 18, 20$ from left to right.
Bottom row: $N=22, 24, 26, 28$ from left to right. 
The GOE denotes the case of the Gaussian orthogonal ensemble, and the GUE denotes the case of the Gaussian unitary ensemble.
\label{SYK2STv1-TF0-N14-28-bare-Ps}}
\end{figure*}

\begin{figure*}
\includegraphics[width=1.\textwidth]{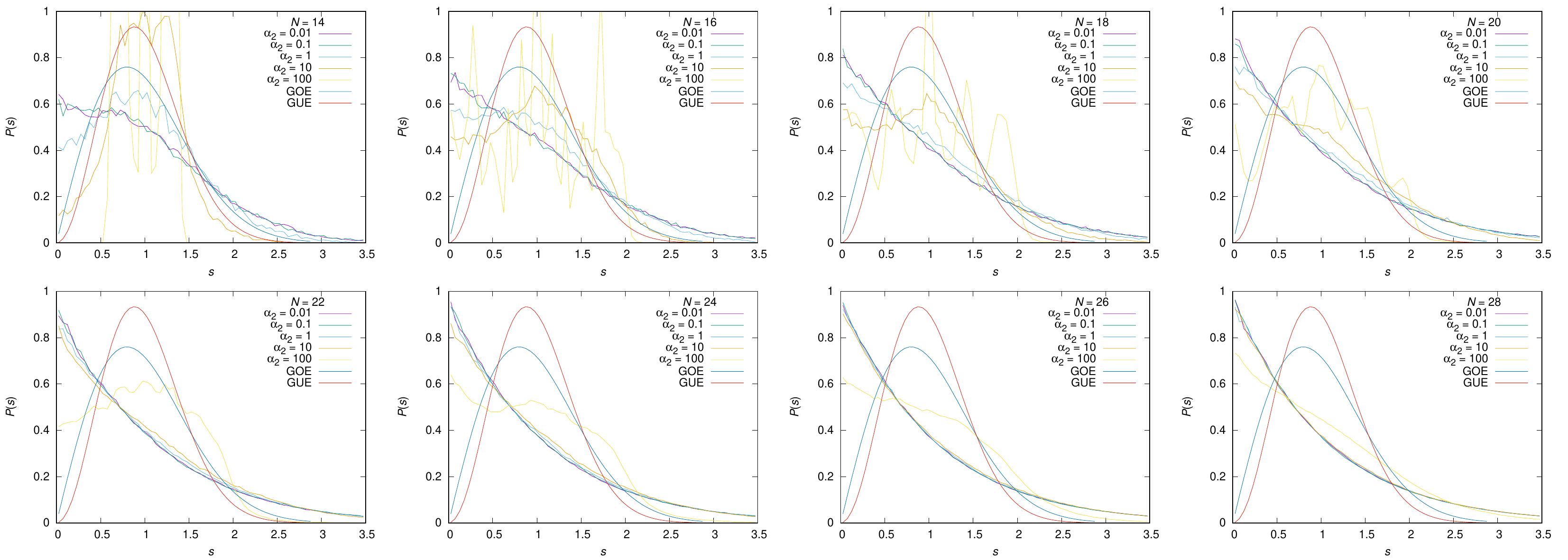}%
\caption{The unfolded generalized level spacing distribution function for the SYK$_2$ model + 2-point source term case ($\alpha_2=0.01, 0.1, 1, 10, 100$).
Top row: $N=14, 16, 18, 20$ from left to right.
Bottom row: $N=22, 24, 26, 28$ from left to right. 
The GOE denotes the case of the Gaussian orthogonal ensemble, and the GUE denotes the case of the Gaussian unitary ensemble.
\label{SYK2STv1-TF0-N14-28Ps}}
\end{figure*}

\begin{figure*}
\includegraphics[width=1.\textwidth]{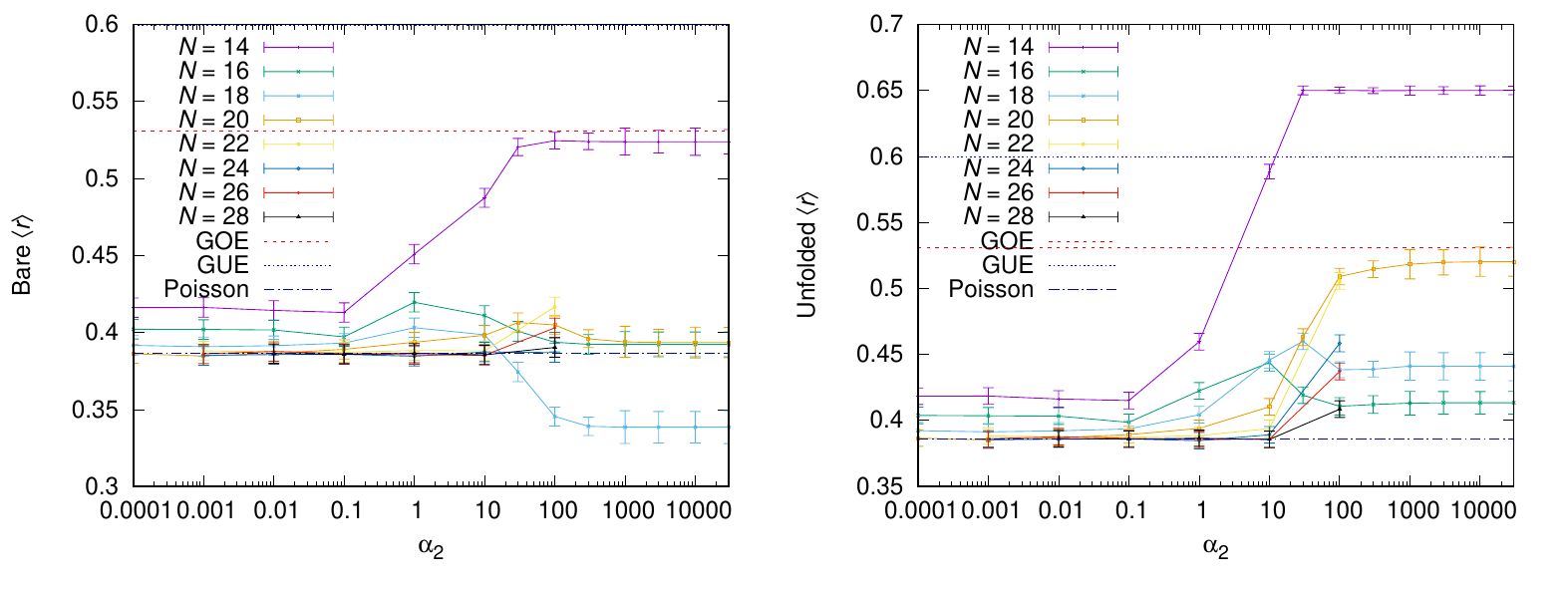}%
\caption{The values of $\langle r \rangle$ for the SYK$_2$ model + 2-point source term case with $N=14, 16, 18, 20, 22, 24, 26, 28$. 
The GOE denotes the case of the Gaussian orthogonal ensemble, the GUE denotes the case of the Gaussian unitary ensemble, and the Poisson denotes the case of the Poisson distribution.
\label{SYK2STv1-T-F0-r-v3}}
\end{figure*}

\begin{figure*}
\centering
\includegraphics[width=0.5\textwidth]{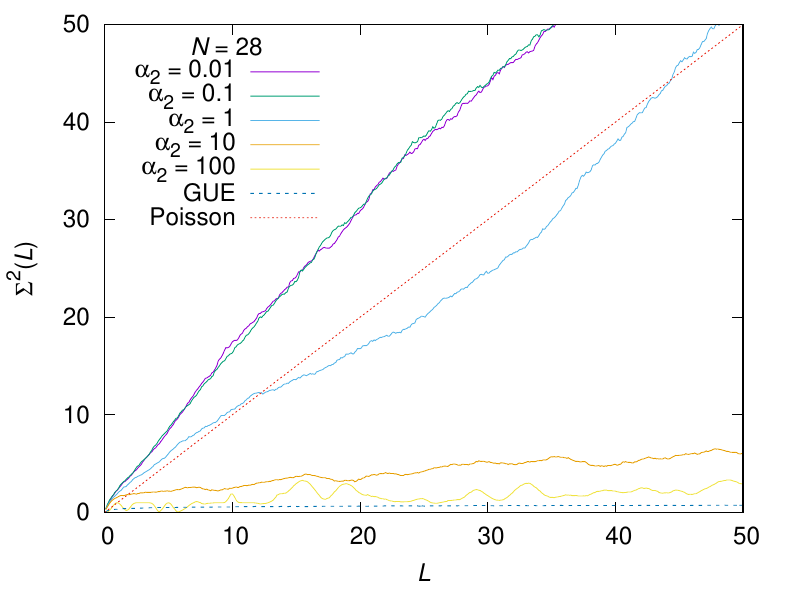}
\caption{The number variance for the SYK$_2$ model + 2-point source term case for $N = 28$.
}
\label{fig:Sigma2-2point}
\end{figure*}

\begin{figure*}
\centering
\includegraphics[width=1.\textwidth]{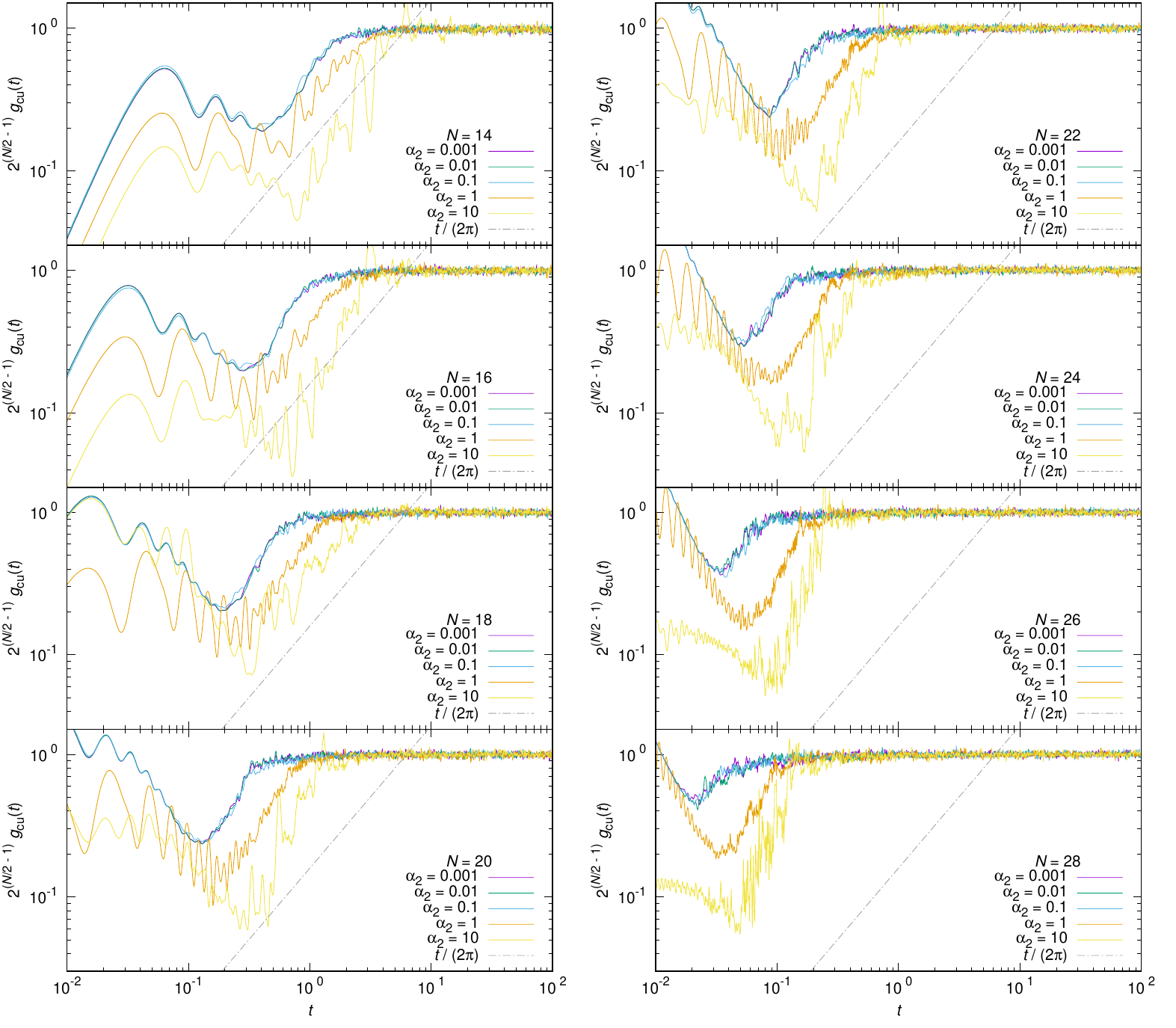}
\caption{The connected part of the unfolded generalized spectral form factor $g_\mathrm{cu}$ for $\beta=0$, $N = 14,16,18,20,22,24,26,28$ and various values of $\alpha_2$. The values of $2^{(N/2)-1}g_\mathrm{cu}$ is shown so that the late-time value is unity.
A line corresponding to $g_\mathrm{cu} = Ct$, with $C = 1/(2\pi D) = 2^{-N/2}/\pi$, \cite{Brezin-Hikami-1996} is also included.
\label{fig:SYK2STv1-TF0-cuSFF}}
\end{figure*}

\noindent
The results of our numerical computations are plotted below. In Figs. \ref{SYK2STv1-TF0-N14-28-bare-Ps} and \ref{SYK2STv1-TF0-N14-28Ps}, we plot the bare and, respectively, unfolded generalized level spacing distribution functions for the two-point source term with coupling values $\alpha_2=0.01, 0.1, 1, 10, 100$ and $N=14, 16, 18, 20, 22, 24, 26$ in the SYK$_2$ model. In Fig. \ref{SYK2STv1-T-F0-r-v3}, we plot $\langle r\rangle$ for the two-point source term case with $N=14, 16, 18, 20, 22, 24, 26, 28$ in the SYK$_2$ model.
Fig.~\ref{fig:Sigma2-2point} shows the number variance $\Sigma^2(L)$. The growth is linear in $L$ and faster than the uncorrelated case for $\alpha_2 \lesssim 1$. The variance is somewhat suppressed for larger $\alpha_2$ and exhibits several peaks rather than simply increase, presumably because if the source term dominates the Hamiltonian, the eigenvalues approach that of the source term and the sample-by-sample fluctuation is localized in the spectrum. Nevertheless, at $\alpha_2 = 100$, for $L\gtrsim 10$, the variance stays larger than that for the GUE.
Fig.~\ref{fig:SYK2STv1-TF0-cuSFF} shows the connected part of the generalized spectral form factor for the unfolded eigenvalues (unfolded spectral form factor for short),
\begin{align}
g_\mathrm{cu}(t, \beta) = \frac{\langle |\tilde{Z}_{\mathrm{SYKg}_2}(\beta, t)|^2\rangle}{\langle |\tilde{Z}_{\mathrm{SYKg}_2}(\beta, 0)|^2\rangle}
- \left(\frac{\langle |\tilde{Z}_{\mathrm{SYKg}_2}(\beta, t)|\rangle}{\langle |\tilde{Z}_{\mathrm{SYKg}_2}(\beta, 0)|\rangle}\right)^2,
\label{eqn:gcut}
\end{align}
in which $\tilde{Z}_{\mathrm{SYKg}_2}(\beta, t)$ is obtained for the entire set of the unfolded eigenvalues $(\tilde{\epsilon}_j)$:
\begin{align}
\tilde{Z}_{\mathrm{SYKg}_2}(\beta, t) = \sum_j \exp(-(\beta + it) \tilde{\epsilon}_j),
\end{align}
for $\beta=0$.
Except at a very early time that reduces exponentially as a function of $N$, the value of $g_\mathrm{cu}(t,\beta=0)$ is flat and does not exhibit a linear increase expected for GUE \cite{Brezin-Hikami-1996}, shown by dash-dotted lines in the plots.
From the above figures, it is clear that turning on a 2-point source term does not generate the random matrix level statistics at large $N$. Consequently, we move on to a 4-point source term.

\subsection{SYK$_2$ Model with 4-Point Source Term}

\begin{figure*}
\centering
\includegraphics[width=0.5\textwidth]{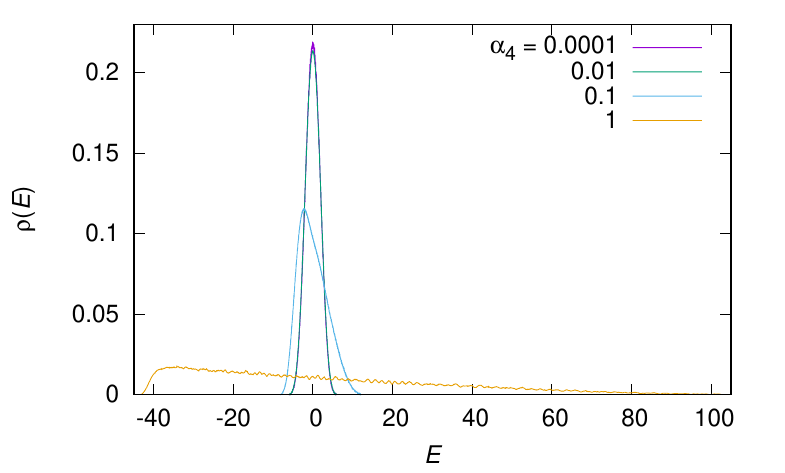}
\caption{
The density of states for the SYK$_2$ model + 4-point source term case with $N = 26$ and $\alpha_4 = 10^{-4}, 10^{-2}, 10^{-1}$ and $1$.
\label{fig:SYK2-4ps-dos}}
\end{figure*}

\begin{figure*}
\includegraphics[width=1.\textwidth]{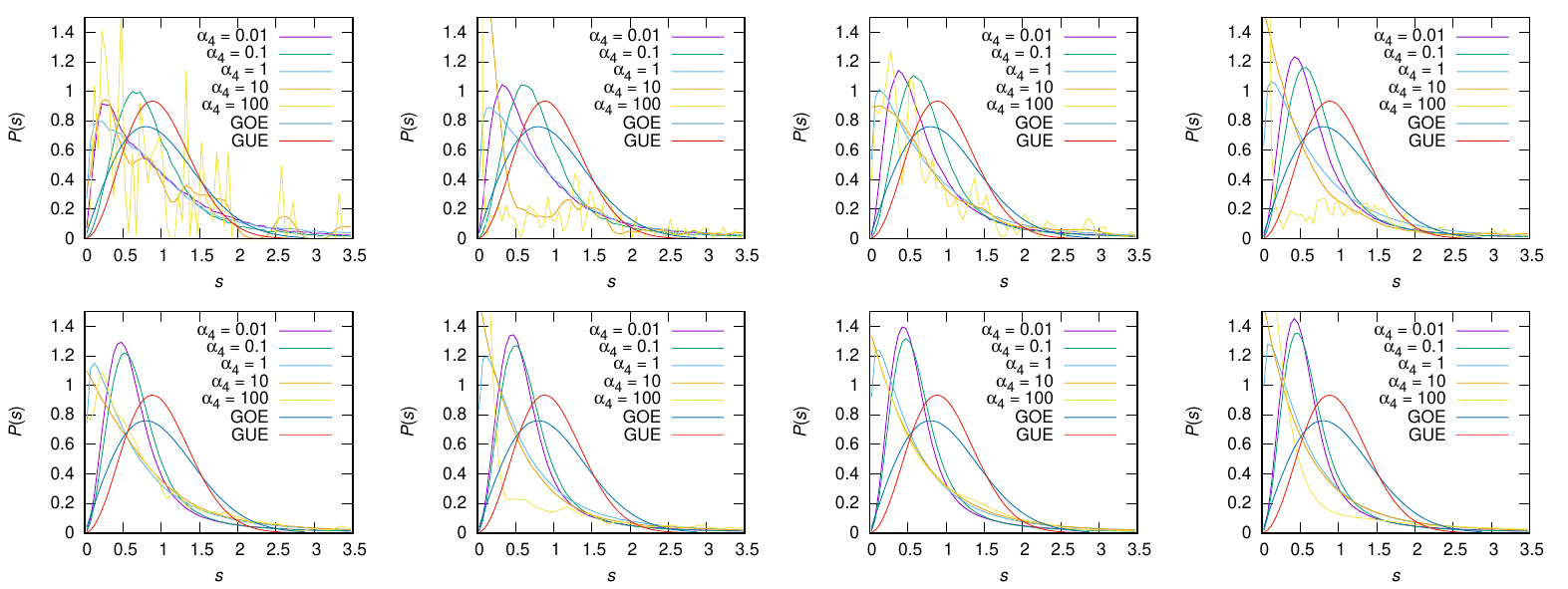}%
\caption{The bare generalized distribution function for the SYK$_2$ model + 4-point source term case ($\alpha_4=0.01, 0.1, 1, 10, 100$).
Top row: $N=14, 16, 18, 20$ from left to right. The distribution does not show the GUE.
Bottom row: $N=22, 24, 26, 28$ from left to right. The GOE denotes the case of the Gaussian orthogonal ensemble, and the GUE denotes the case of the Gaussian unitary ensemble.
\label{SYK-bare-Ps-N14-28}}
\end{figure*}

\begin{figure*}
\includegraphics[width=1.\textwidth]{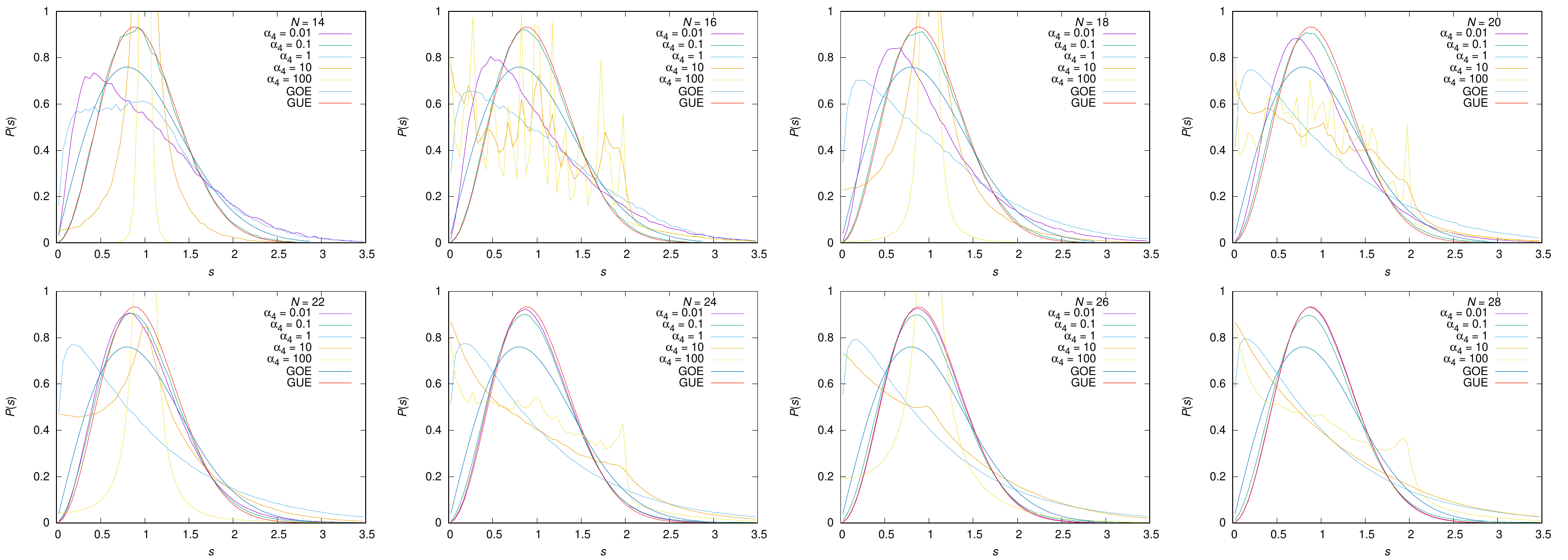}
\caption{The unfolded generalized distribution function for the SYK$_2$ model + 4-point source term case ($\alpha_4=0.01, 0.1, 1, 10, 100$).
Top row: $N=14, 16, 18, 20$ from left to right. When $N$ goes to $24, 26, 28$, the distribution shows the GUE. This gives a different result compared to the bare distribution.
Bottom row: $N=22, 24, 26, 28$ from left to right. The GOE denotes the case of the Gaussian orthogonal ensemble, and the GUE denotes the case of the Gaussian unitary ensemble.
\label{SYK2STv1-T0F-N14-28Ps}}
\end{figure*}

\begin{figure*}
\includegraphics[width=1.\textwidth]{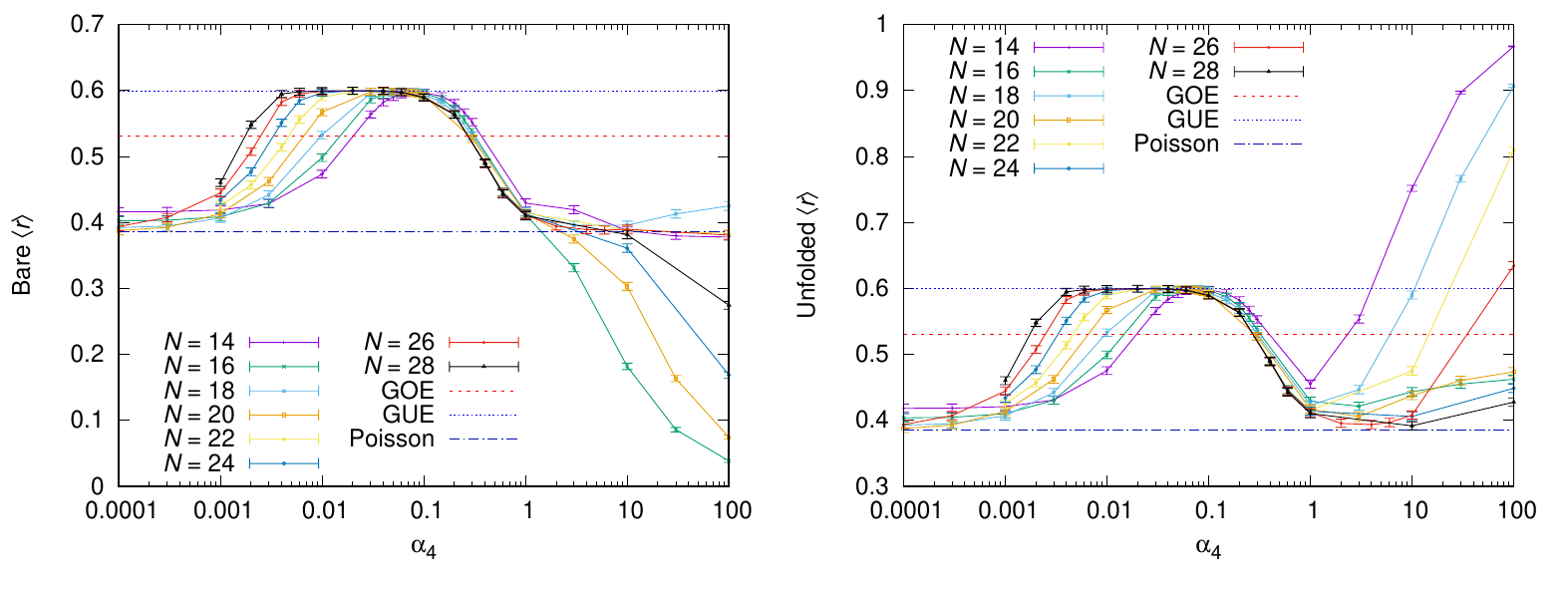}%
\caption{The values of $\langle r \rangle$ for the SYK$_2$ model + 4-point source term case from $N=14, 16, 18, 20, 22, 24, 26, 28$. We find the GUE in the perturbation region (at around $\alpha_4=0.01-0.1$), but the bare result is not self-consistent between $\langle r\rangle$ and $P(s)$.
The GOE denotes the case of the Gaussian orthogonal ensemble, the GUE denotes the case of the Gaussian unitary ensemble, and the Poisson denotes the case of the Poisson distribution.
\label{SYK2STv1-T0F-r}}
\end{figure*}

\begin{figure*}
\includegraphics[width=1. \textwidth]{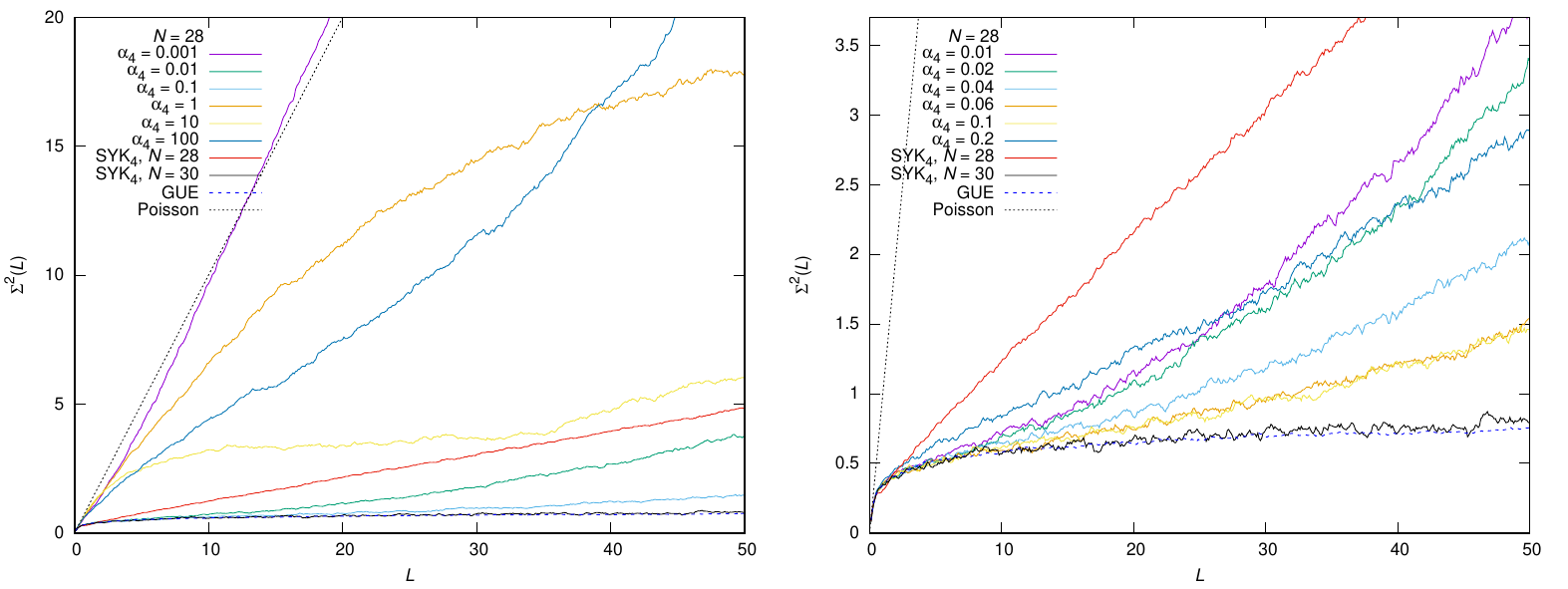}
\caption{The number variance for the SYK$_2$ model + 4-point source term case.
Left: $\alpha_4=0.001,0.01,0.1,1,10,100$. Right: $\alpha_4=0.01,0.02,0.04,0.06,0.1,0.2$.
The data for the SYK$_4$ model with $N = 28$ (3000 samples) and $N = 30$ (914 samples) are also shown.
Note that the SYK$_4$ model with $N = 28$ shows GOE behavior while the SYK$_2$ model + 4-point source term shows GUE behavior even for $N = 28$. 
The GUE denotes the case of the Gaussian unitary ensemble, and the
Poisson denotes the case of the Poisson distribution. 
\label{fig:Sigma2-4point}}
\end{figure*}

\begin{figure*}
\includegraphics[width=0.5\textwidth]{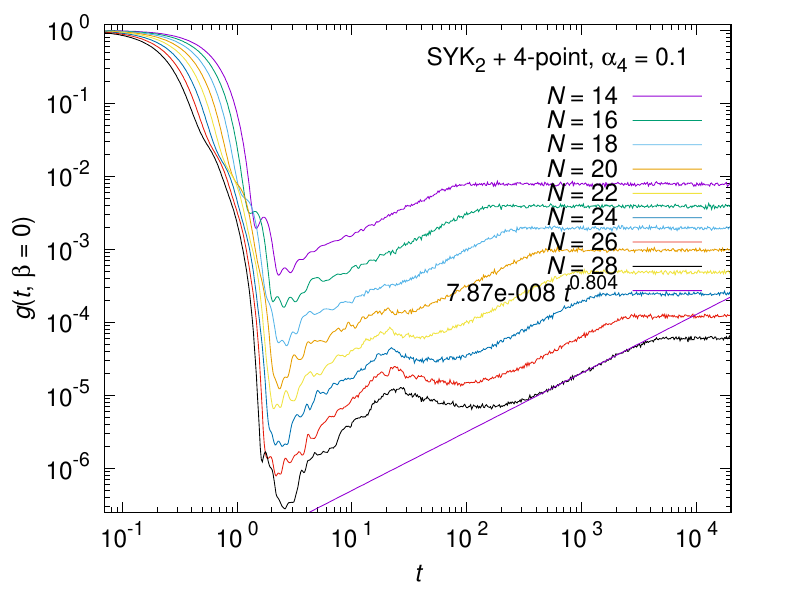}
\includegraphics[width=0.5\textwidth]{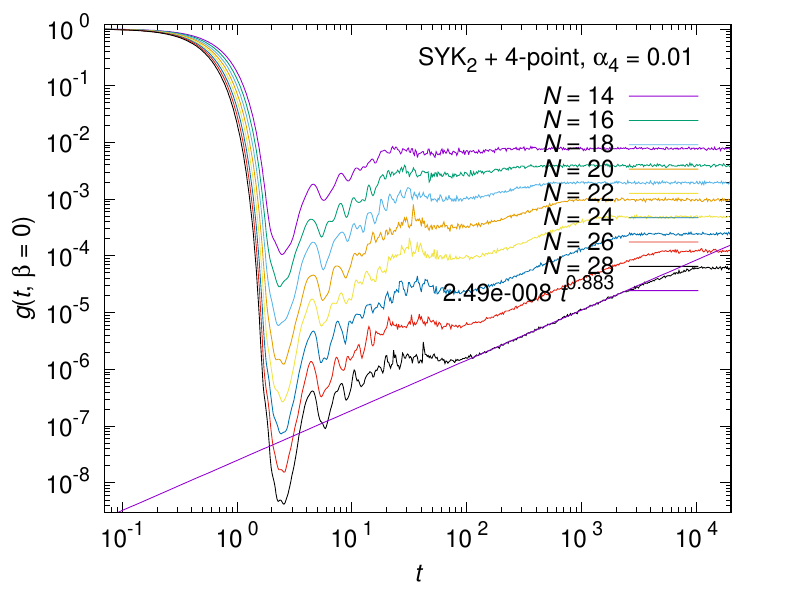}
\includegraphics[width=0.5\textwidth]{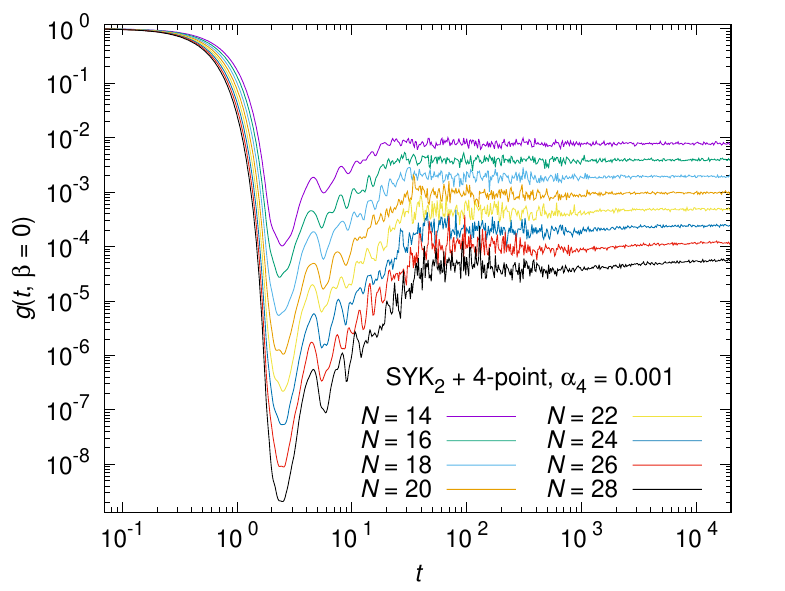}
\caption{The generalized spectral form factor for the SYK$_2$ model+4-point source term case. 
Top row: $\alpha_4 = 0.1$ (left) and $\alpha_4 = 0.01$ (right). 
Bottom row: $\alpha_4=0.001$. The figures with $\alpha_4=0.1$ and $\alpha_4=0.01$ show the level statistics of the random matrix (GUE) but $\alpha_4=0.001$ does not. The comparison exhibits a feature that the spectral form factor has a smooth linear growth for the GUE.
\label{SYK2STv1-T0F-N-Gt}}
\end{figure*}

\begin{figure*}
\centering
\includegraphics[width=1.\textwidth]{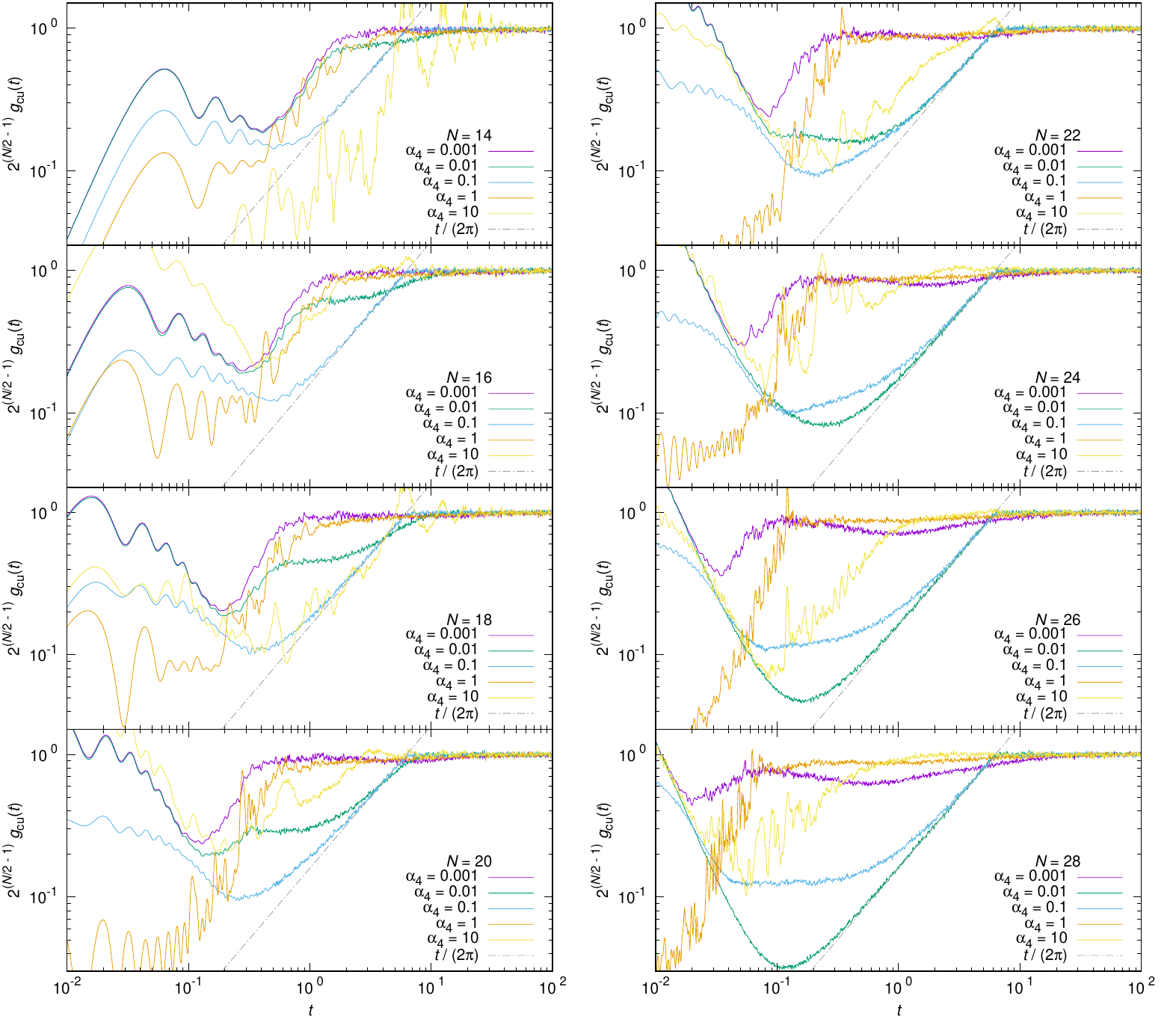}
\includegraphics[]{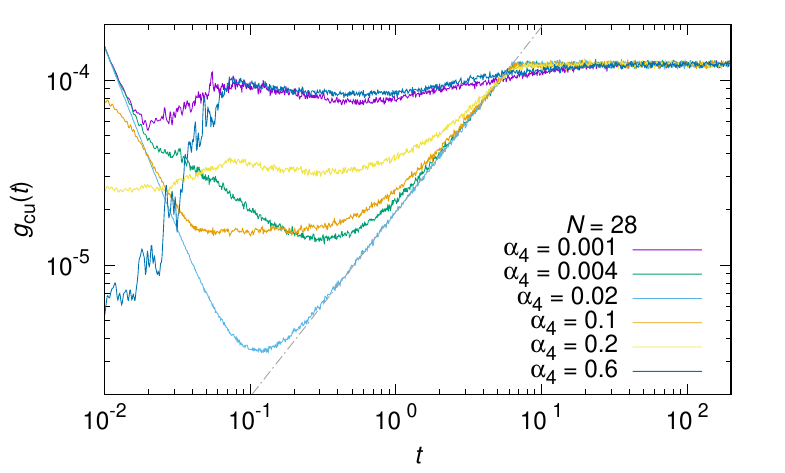}
\caption{Upper: The connected part of the unfolded generalized spectral form factor $g_\mathrm{cu}(t,\beta=0)$ for $N = 14,16,18,20,22,24,26,28$ and various values of $\alpha_4$. The values of $2^{(N/2)-1}g_\mathrm{cu}$ is shown so that the late-time value is unity.
A dash-dotted line corresponding to $g_\mathrm{cu} = Ct$, with $C = 1/(2\pi D) = 2^{-N/2}/\pi$, Ref. \cite{Brezin-Hikami-1996} is also included.
Lower: The bare value of $g_\mathrm{cu}(t,\beta=0)$ for $N = 28$ for a set of values of $\alpha_4$ different from the upper panel.
\label{fig:SYK2STv1-T0F-cuSFF}}
\end{figure*}

\noindent
We summarize the results of our numerical computations here:
\begin{itemize}
   \item The density distribution $\rho(E)$ of energy eigenvalues for various strengths of the 4-point source term is shown in Fig.~\ref{fig:SYK2-4ps-dos}. As $\alpha_4$ is increased, the shape of $\rho(E)$ becomes less asymmetric about $E=0$ and approaches the one for the case with only the source term as shown in the left part of Fig.~\ref{fig:source4-dos-gt}.
   \item In Fig. \ref{SYK-bare-Ps-N14-28}, we plot the bare generalized distribution of the energy gap with a 4-point source term turned on for various values of the 4-point coupling. The 4-point couplings are chosen to be $\alpha_4=0.01, 0.1, 1, 10, 100$ and $N=14, 16, 18, 20, 22, 24, 26, 28$. Note that the distribution still does not show the GUE at the large $N$.
   \item In Fig. \ref{SYK2STv1-T0F-N14-28Ps}, we plot the unfolded generalized distribution function for the 4-point source term for the same values of $\alpha_{4}$ and $N$. We note that now the distribution does show GUE behavior for large $N$ for $0.01\lesssim \alpha_4\lesssim 0.1$, giving a qualitatively different result from the bare distribution.
   \item In Fig. \ref{SYK2STv1-T0F-r}, we plot $\langle r\rangle$ for the 4-point source term and $N=14-28$ in the SYK$_2$ model. The GUE is found in the perturbation region for $\alpha_4$ in the range $0.01-0.1$. Curiously, the GUE appears in the bare result for $\langle r\rangle$ but not in the bare distribution, pointing to an inconsistency in the use of the bare spectrum in analyzing the level statistics. The underlying reason for this is not yet clear to us.
   \item Fig.~\ref{fig:Sigma2-4point} shows the number variance $\Sigma^2(L)$. The growth is similar to the uncorrelated (Poisson) case but is significantly suppressed for $0.01\lesssim \alpha_4 \lesssim 0.2$ which coincides at smaller $L$ with the number variance for the GUE ensemble as well as with the SYK$_4$ model with $N = 30$ having the GUE universality. 
   The difference in the large $L$ is expected to be suppressed by using a larger $N$. 
   For larger $\alpha_4$, the number variance is significantly larger and no longer resembles the random matrix results. We use 3000 and 914 samples in the SYK$_4$ model for $N=28$ (with GSE universality) and $N=30$ (GUE) respectively for comparison.
   \item Fig. \ref{SYK2STv1-T0F-N-Gt} shows the generalized spectral form factor in the SYK$_2$ model with the 4-point source term turned on, which we have already shown in Fig.~\ref{SYK2STv1-T0F-F-Gt} using separate plots for each $N$, for $N = 14, 16, 18, 20, 22, 24, 26, 28$ and $\alpha_4=0.001, 0.01, 0.1$. We compare the figures with $\alpha_4=0.1$ and $\alpha_4=0.01$, which exhibits a GUE, to the figure with $\alpha_4=0.001$, which displays no apparent random matrix level statistics. The spectral form factor for $\alpha_4=0.1$ and $\alpha_4=0.01$ shows a smooth linear growth before saturation.
   \item 
Fig.~\ref{fig:SYK2STv1-T0F-cuSFF} shows the connected part of the generalized spectral form factor for the unfolded eigenvalues (unfolded spectral form factor for short), Eq.~\eqref{eqn:gcut}.
\\

\noindent
We observe that the ramp part agrees better to a linear increase compared to the generalized spectral form factor for the bare spectrum, see Fig.~\ref{SYK2STv1-T0F-F-Gt} and Fig.~\ref{SYK2STv1-T0F-N-Gt}. The ramp starts at smaller $t$ as $\alpha_4$ is increased for $0.004\lesssim t \lesssim 0.01$, but for larger $\alpha_4$ the ramp becomes shorter again. For $\alpha_4 \lesssim 0.001$ or $\alpha_4 \gtrsim 0.6$, the linear ramp is not observed, which is consistent with the behavior of the averaged gap ratio $\langle r\rangle$ in Fig.~\ref{SYK2STv1-T0F-r}.
\end{itemize}

\noindent
To summarize: The GUE appears in the perturbation region at around $\alpha_4=0.01-0.1$. Because the correlation function follows Wick's theorem in the SYK$_2$ model, we appear to be finding random matrix level statistics in the near-integrable region of the model. When $\alpha_4$ vanishes, the theory is integrable, and its eigenvalue spectrum obeys a Poisson distribution as expected. When the source coupling is non-zero and in the range $\alpha_4=0.01-0.1$ which is still within the near-integrable region, the Poisson distribution deforms to a GUE. The observation is interesting because the 4-point source term has an integrable spectrum in the large-$N$ limit. This observation is supported by a comparison of the generalized spectral form factor with random matrix level statistics to other values of $\alpha_4$, where the former clearly shows a smooth linear growth before saturation. When $N$ becomes bigger approaching the  GUE, we expect the system to spend longer in the ramp region. This offers a classification for the ramp region: (1) No smooth linear growth corresponds to no GUE; (2) partial smooth linear growth corresponds to mixed behavior, and (3) full smooth linear growth corresponds to GUE behavior. We also compare the result of the computation of $\langle r\rangle$ to $P(s)$ which appears to point to an internal inconsistency in using the bare spectrum as opposed to the unfolded spectrum to analyze the level statistics. 
The linear growth in the ramp region can be understood from the repulsion between the eigenvalues confined around zero that gives a rigid structure. 
This linear ramp is observed both for the spectral form factor and the connected part of the unfolded spectral form factor. 
Therefore, ramp behavior is also generated by the Gaussian random variable in the integrable spectrum. 

\section{Discussion and Conclusion}
\label{sec:4}
\noindent
The introduction of (short- and long-range) correlations in disordered interactions is expected to be consequential to an understanding of (many-body Anderson) localization-delocalization transitions and therefore of some importance in random many-body physics. In this article, we studied this question in the context of the SYK$_{2}$. Specifically, we computed the generalized spectral form factor and level spacing distribution in the SYK$_2$ model including the information of 2- and 4-point correlations into the level statistics. For the case of $N=2$ Majorana fermions, the generalized spectral form factor can be computed exactly. This exact solution captures the essential behavior of the larger $N$ cases with a source term and offers an interpretation of the general dip-ramp-plateau behavior. When we include a 4-point source term, the generalized level spacing distribution, and the mean value of the adjacent gap ratio show GUE \cite{Dyson:1962es} for perturbatively small values of $\alpha_{4}$. 
\\

\noindent
In 1977, Berry conjectured the relation between the eigenstates of a system and its non-integrability properties \cite{Berry:1977zz}; specifically, a non-integrable system should exhibit Gaussian random eigenstates. The question, however, is: How to diagnose this best in such a system when the Hamiltonian cannot be exactly diagonalized? One option lies with the spectral form factor. The SYK$_2$ model, for example, with its quadratic interaction is expected to be integrable. However, except for the very special $N=2$ case,  we find that the model can also exhibit Gaussian random eigenstates. This result is supported by the spectral form factor which displays a dip-ramp-plateau behavior for $N>2$ \cite{Lau:2018kpa}. It would seem then that this dip-ramp-plateau behavior answers the question. However, we also find that even for $N=2$, for sufficiently large $\alpha_2$, even the exact spectral form factor displays dip-ramp-plateau behavior. At the level of the exact SFF, the reason for this behavior can be attributed to the fixed and non-zero coefficients of the source term. The source term suppresses the randomness of the system when $\alpha_2$ is sufficiently large. Again, this gels with the idea that correlations in the random couplings can potentially induce localization-delocalization transitions in many-body physics \cite{Monteiro:2020buv}.
\\

\noindent
We also studied both the bare and unfolded level spacing distribution and found that only when a change of the gap average is slow, both level spacing distributions give a qualitatively similar result. Here our results show that only the unfolded distribution is self-consistent by comparing $P(s)$ to $\langle r\rangle$. This agrees with the popular belief that the unfolded level spacing distribution is more suitable for a study of the level statistics. As was proven in a generic bosonic quantum mechanical system from the spectral form factor \cite{Muller:2004nb}, a chaotic system should exhibit random matrix level statistics \cite{Bohigas:1983er}. However, because this proof hinges on the WKB approximation, it is strictly only valid for a short time. A full-time result is essential to understand the relation between quantum chaos and the random matrix rigorously. Our results show the emergence of random matrix theory in a certain near-integrable region of the deformed SYK$_{2}$ model but deviate from this behavior when the coupling constant becomes larger in contrast with any expectation that the chaos grows with strengthened coupling. In this paper, we considered a fermionic system. The fermionic fields necessarily satisfy an anti-commutation relation instead of the commutation relation of the bosonic fields. Consequently, taking the classical limit in the fermionic system is more subtle. Therefore, our results should not be taken as a contradiction of the proof in Ref. \cite{Muller:2004nb}, so much as a loophole.
\\

\noindent
In a numerical study, it is difficult to demonstrate exact random matrix behavior at a finite $N$. Therefore, understanding the difference between the random and non-random matrix behavior is of some importance. With a 4-point source term turned on, we found that the smooth linear growth in the ramp region can indeed distinguish GUE from non-random matrix behavior so it would be of great interest to study $1/N$ corrections in this system.
\\

\noindent
Finally, we would like to comment on the random matrix behavior of this system. 
Because the SYK$_2$ model and 4-point source terms both provide integrable spectra, as diagnosed from the level spacing distribution function, any random matrix behavior appears only in the intermediate region of the source terms, but not in the strong or weak source term regimes. 
While a similar phenomenon first appeared in Ref. \cite{Friedrich:1989tf},
because the fermionic system does not have a clear classical limit, our findings in the SYK-type model is novel and interesting. However, the appearance of the random matrix behavior in the level statistics implies that the long time dynamics of a system is insensitive to the initial state, but this only constrains the late time behavior of the system. On the other hand, the Lyapunov exponent quantifies the sensitivity of the early-time dynamics to the initial conditions in finite-dimensional quantum mechanical systems. In this sense, it is left for future study to characterize the chaotic regime between integrable limits of such models in terms of positivity of the Lyapunov exponent.

\section*{Acknowledgments}
\noindent
We would like to thank Antonio M. García-García, Paolo Glorioso, Masanori Hanada, Aitor Lewkowycz, Wolfgang Mück, Laimei Nie, Xiao-Liang Qi, Dario Rosa, Shinsei Ryu, and Stephen H. Shenker for useful discussions. CTM would like to thank Nan-Peng Ma for his encouragement.
\\

\noindent
PHCL was supported by the postdoctoral fellowship of the National Center for Theoretical Sciences. 
CTM was supported by the Post-Doctoral International Exchange Program; 
China Postdoctoral Science Foundation, Postdoctoral General Funding: Second Class (Grant No. 2019M652926); 
Foreign Young Talents Program (Grant No. QN20200230017); 
Science and Technology Program of Guangzhou (Grant No. 2019050001). JM was supported in part by the NRF of South Africa under grant CSUR 114599. 
MT was partially supported by Grants-in-Aid No.~JP17K17822, No.~JP20H05270, and No.~JP20K03787 from JSPS of Japan. 
\\

\noindent
We would like to thank the National Tsing Hua University, Kadanoff Center for Theoretical Physics, Shing-Tung Yau Center at the Southeast University, and Institute of Theoretical Physics at the Chinese Academy of Sciences, for hospitality at various stages of this work.

\end{document}